\documentclass[journal]{vgtc}                

\ifpdf
  \pdfoutput=1\relax                   
  \usepackage{graphicx}                
  \DeclareGraphicsExtensions{.pdf,.png,.jpg,.jpeg} 
\else
  \usepackage{graphicx}                
  \DeclareGraphicsExtensions{.eps}     
\fi%

\graphicspath{{figs/}{pictures/}{images/}{./}} 

\usepackage{microtype}                 
\PassOptionsToPackage{warn}{textcomp}  
\usepackage{textcomp}                  
\usepackage{mathptmx}                  
\usepackage{times}                     
\usepackage{cite}                      
\usepackage{tabu}                      
\usepackage{booktabs}                  

\usepackage{review}
\usepackage{xspace}
\usepackage{subfigure}

\usepackage[utf8]{inputenc}
\usepackage{xcolor}
\usepackage{booktabs} 
\usepackage{textcomp} 

\usepackage{hyperref}

\usepackage{fontawesome}

\makeatletter
\usepackage{enumitem}
\setenumerate{noitemsep,topsep=2pt,parsep=0pt,partopsep=2pt}
\setitemize{noitemsep,topsep=2pt,parsep=0pt,partopsep=2pt}
\usepackage{stfloats}

\onlineid{0}

\vgtccategory{Research}
\vgtcpapertype{algorithm/technique}

\newcommand{\house}{\faRebel}
\newcommand{\CHI}{\faStreetView}
\newcommand{\person}{\faUser}
\newcommand{\TVCG}{\faAreaChart}

\newcommand{\battle}{\faCrosshairs }

\title{Juniper: A Tree+Table Approach to Multivariate Graph Visualization}

\author{Carolina Nobre, Marc Streit, and Alexander Lex}
\authorfooter{
\item
 Carolina Nobre is with the University of Utah. E-mail: cnobre@sci.utah.edu.
\item
 Marc Streit is with Johannes Kepler University Linz. E-mail: marc.streit@jku.at.
\item
 Alexander Lex is with the University of Utah. E-mail: alex@sci.utah.edu.
}

\shortauthortitle{Nobre \MakeLowercase{\textit{et al.}}: Juniper: A Tree+Table Approach to Multivariate Graph Visualization}

\abstract{
Analyzing large, multivariate graphs is an important problem in many domains, yet such graphs are challenging to visualize. In this paper, we introduce a novel, scalable, tree+table multivariate graph visualization technique, which makes many tasks related to multivariate graph analysis easier to achieve. The core principle we follow is to selectively query for nodes or subgraphs of interest and visualize these subgraphs as a spanning tree of the graph. The tree is laid out linearly, which enables us to juxtapose the nodes with a table visualization where diverse attributes can be shown. We also use this table as an adjacency matrix, so that the resulting technique is a hybrid node-link/adjacency matrix technique. We implement this concept in Juniper and complement it with a set of interaction techniques that enable analysts to dynamically grow, restructure, and aggregate the tree, as well as change the layout or show paths between nodes. 
We demonstrate the utility of our tool in usage scenarios for different multivariate networks: a bipartite network of scholars, papers, and citation metrics and a  multitype network of story characters, places, books, etc.
} 


\CCScatlist{ 
 \CCScat{K.6.1}{Management of Computing and Information Systems}%
{Project and People Management}{Life Cycle};
 \CCScat{K.7.m}{The Computing Profession}{Miscellaneous}{Ethics}
}

\teaser{
  \centering
  \includegraphics[width=\linewidth]{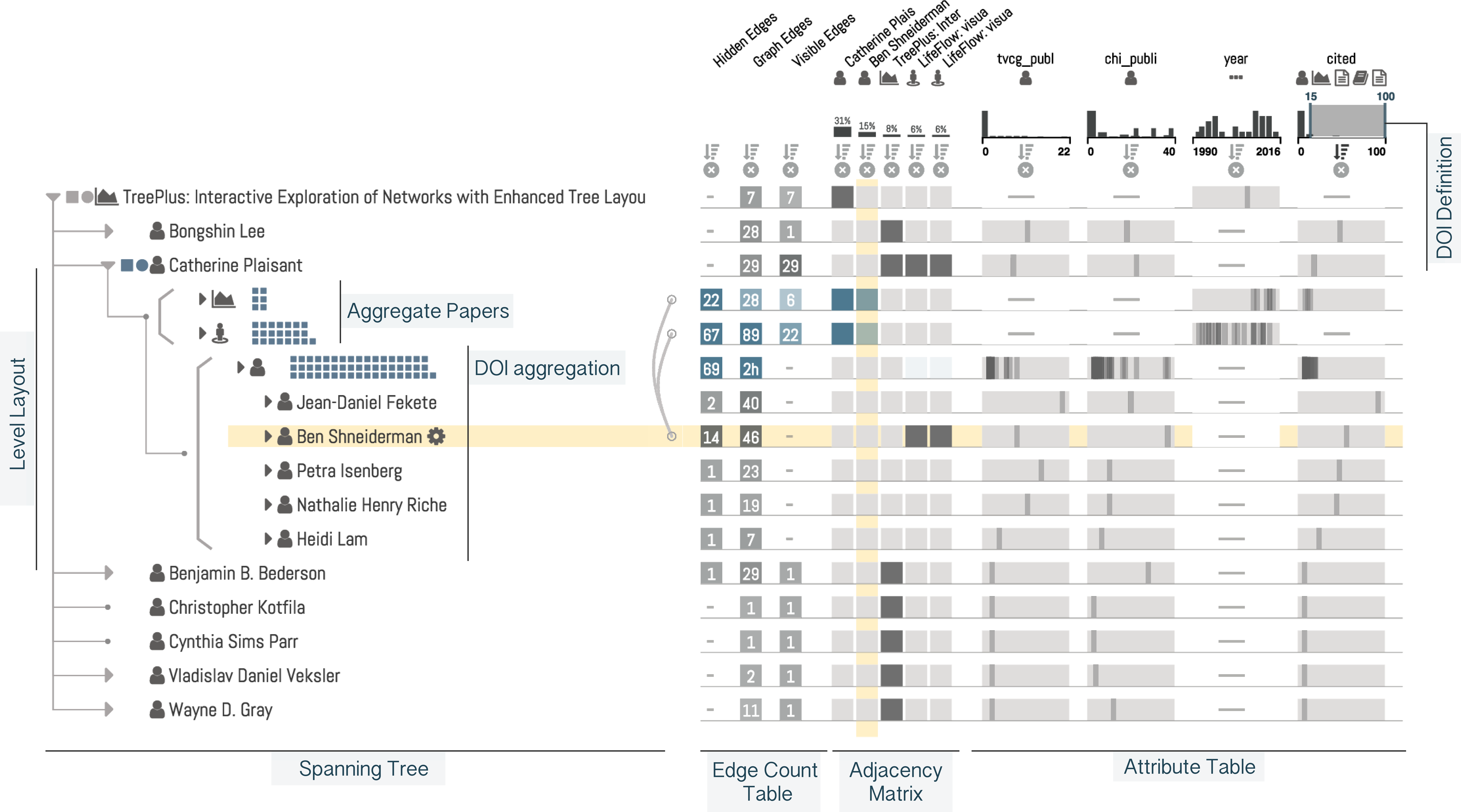}
  \vspace{-7mm}
  \caption{Juniper visualizing a co-author network starting at the TreePlus paper as a spanning tree. The graph is extended for Catherine Plaisant to include all her papers and co-authors. The papers are shown in aggregate form and faceted by \CHI~CHI and \TVCG~TVCG. Most of the tree use a conventional layout, but the descendants of Catherine Plaisant's node are shown in level layout, which groups nodes by distance to the branch root. Nodes in this branch are aggregated, with the exception of prolific authors, which are revealed using a degree-of-interest function. Ben Shneiderman is highlighted; two hidden edges originate at his node. The edge-count table shows a summary of the connectivity of each node. The adjacency matrix shows explicit connections to selected, highly connected nodes. The attribute table shows attributes about the authors and papers for individual as well as aggregated rows.}
	\label{fig:juniper_overview}
	  \vspace{-2mm}
}

\renewcommand{\manuscriptnotetxt}{}

\begin{document}


\firstsection{Introduction}


\maketitle

Graph visualization is a challenging problem, especially when the size of the graph exceeds a few hundred nodes. This lack of scalability is exacerbated when rich attributes for the nodes and/or the links need to be considered when analyzing a graph. Such multivariate graphs are common across domains: biologists, for example, need to explore canonical pathways in the context of experimental data, to judge whether a pathway is valid for a given tissue or organism; social scientists may need to study whether a tight group of friends are all in the same age group and went to the same school. The difficulty of visualizing multivariate networks arises from two conflicting goals that need to be reconciled: visualizing topology and visualizing node and edge attributes. The visualization community has a good understanding of how to visualize either the topology of a network or the multidimensional data that is associated with the nodes and edges, yet addressing both topology-based tasks and attribute-based tasks at the same time is still an open research problem.  While there has been progress on visualizing aggregate attributes for the larger structure of a graph~\cite{vandenelzen_multivariate_2014} or on visualizing attributes for special graph structures such as trees~\cite{nobre_lineage:_2018} or paths~\cite{partl_enroute:_2012, partl_pathfinder:_2016}, we are not aware of a scalable, multivariate graph visualization technique that excels at supporting \textit{focus tasks}. Here, we introduce such a technique.  

We use the term focus tasks to refer to tasks where the details of individual nodes, edges, and their neighborhood matter, as opposed to the global structure of the network. Focus tasks commonly require readable labels and a detailed understanding of a node's attributes. These tasks include identifying adjacent nodes (who are my friends?), identifying nodes that are accessible from another node (where can I fly to from this airport within at most one layover?), finding short paths (what's the best route to go from A to B?), etc. Examples for focus graph tasks on multivariate networks include investigating congestion and latency in a computer network or exploring how a mutated gene influences activity levels of the genes in its neighborhood. It is worth noting that these focus graph tasks are equally important in both large and small graphs.


Our primary contribution is Juniper, a new interactive technique that is tailored to address focus tasks when visualizing large, multivariate networks. The core idea is to extract a spanning tree from a subgraph that is the result of a query of a larger graph. The spanning tree is grown from a node of interest and laid out in a linearized tree, where every node can be unambiguously associated with a row in a table. This table is used to visualize topological properties of the tree, such as the degree of the nodes and their adjacency to selected other nodes, and to show rich attributes. 

We also contribute an implementation of this technique, which enriches this basic concept with user interactions to restructure the tree to best answer the analyst's question, expose additional topological information such as edges not included in the tree, identify shortest paths between nodes, explore interdependent attributes along paths in the network, aggregate groups of nodes to save space, expand the network on demand, filter nodes by type, or sort them based on attributes.


Juniper is tailored to address focus tasks related to the details of a large network.
We argue that this class of tasks is important in many practical applications and complementary to overview tasks that are better addressed with other techniques. 

\section{Data and Tasks}
\label{sec:tasks}

We consider graphs $G = (N, E)$ with nodes $n \in N$ and edges $e \in E$, which can be of different types $t \in T$. Edges can be directed. Nodes have attributes $a \in A_{nodes}$ associated with them. Typically, nodes of different types also have different attributes. Node attributes can be numerical, ordinal, nominal, sets, or labels/identifiers. Although our prototype does not currently support it, conceptually we could also incorporate edge attributes. 
Juniper renders a subset of the graph $g_{sub} \in G$, where $|g_{sub}|\ll|G|$. This subgraph is selected by an analyst to satisfy a specific question and can change over the course of an analysis. Subgraphs do not have to be connected.

Whereas many graph visualization techniques are designed to support overview tasks and to be scalable with respect to the absolute number of nodes, edges, and attributes shown, only a few graph visualization tasks require getting a large-scale overview of a network. We consider all tasks where analysts need to see a large set of nodes and edges to be \textit{overview tasks}. Examples of such overview tasks are estimating the size of a network, identifying clusters, or finding articulation points. An example for multivariate networks is to explore how migration patterns within the US differ by age. When visualizing overviews of all but trivial networks, the large number of nodes and edges makes it impossible to show labels and attributes for individual nodes.
For \textbf{focus tasks}, the details of a small, well-defined subset of nodes are relevant and necessary for the task. These details include topological information such as neighborhoods of or paths between nodes; and attributes, including node labels and other associated data.

 
To get a better sense of the importance of focus tasks, we classified Lee et al.'s task taxonomy for graph visualization~\cite{lee_task_2006} into whether the tasks are focus tasks or require an overview. Of nine tasks Lee et al. identify, five are focus tasks (adjacency, accessibility, common connection, follow path, revisit). The attribute-related tasks --- node attributes, link attributes --- are described mostly in a focus context by Lee et al., but they can also be useful in a global context (for example, estimating the average age of members of a social network). One task --- connectivity --- can be broken up into overview and focus tasks. For example, finding the shortest path between nodes is a focus connectivity task, but identifying clusters, connected components, bridges, or articulation points is an overview variant of the connectivity task.

With regard to topology-attribute interaction, focus tasks can be classified into two groups: (1) those that can be achieved in the context of neighborhoods (adjacency, accessibility, common connection), e.g., to see whether friends have similar educational attainment or whether health issues, such as obesity, spread in a neighborhood of friends~\cite{christakis_spread_2007}, and (2) those related to exploring attributes in the context of paths (follow path, connectivity), for example, to judge delays over time in a computer network or whether a path in a biological pathway is active in a set of samples~\cite{partl_enroute:_2012}.

Juniper is designed to support focus tasks, specifically the types of tasks that are concerned with both topology and attributes. We employ a bottom-up graph visualization technique~\cite{vanham_search_2009, vonlandesberger_visual_2011} where analysts start with a query and expand the network on demand. As such, it is well suited to answer questions about specific subnetworks, but it cannot give large-scale overviews of the network.

\section{Related Work}

Juniper is inspired by and contributes to multiple subfields of graph visualization. Here we discuss how our work relates to multivariate graph and tree visualization, to tree-based graph visualization, and to query-based visualization of large graphs.

\subsection{Multivariate Graph and Tree Visualization} 

A multivariate graph is a graph where the nodes and/or edges are associated with attributes~\cite{kerren_multivariate_2014}. Although most graphs have some attributes, such as a node type, multivariate graph visualization techniques are concerned with graphs with several or even hundreds of associated attributes. A common goal of multivariate graph visualization techniques is to allow analysts to jointly analyze topology and attributes and reason about their relationship. Partl et al.~\cite{partl_enroute:_2012} discuss four different types of multivariate graph visualization techniques, based on node-link layouts, which we use to structure this section. We also discuss matrix-based techniques as a fifth type.

\noindent \textbf{(1) On-node encoding} refers to modifying the visual appearance of a node (size, color), or embedding marks in it (bar charts, line charts, etc.) Color coding is a common choice to encode a single data value or a node type; the latter is also often encoded using node shapes or icons. Gehlenborg et al.~\cite{gehlenborg_visualization_2010} review techniques used in systems biology for visualizing multivariate networks, many of which make use of on-node encoding using embedded charts, such as line charts, box plots, etc. On-node encoding is also widely supported by common graph visualization tools such as Cytoscape~\cite{smoot_cytoscape_2011} and Gephi~\cite{bastian_gephi:_2009}. Van den Elzen and van Wijk~\cite{vandenelzen_multivariate_2014} use embedded visualizations to show distributions of values aggregated in a super-node. On-node encoding supports the integration of topology and attribute-based tasks well; however, it comes with scalability trade-offs. Even for a modest number of nodes in a node-link layout, node size has to be limited; hence little space is available to encode attributes. When details about nodes are shown, as, for example, in MoireGraphs~\cite{jankun-kelly_moiregraphs:_2003}, the number of nodes that can be displayed simultaneously is limited.

\noindent \textbf{(2) Multiple coordinated view (MCV)} approaches use separate, dedicated views for the attributes and the topology. Common examples are combinations of force-directed node-link diagrams with multidimensional data visualization techniques~\cite{shannon_multivariate_2008, lex_caleydo:_2010}, or providing a detail view for individual nodes~\cite{heer_vizster:_2005, tu_graphcharter:_2013}. Although this solution is flexible and easy to implement, it requires interactive highlighting to identify relationships between nodes and their attributes. MCV-based attribute visualization is supported by standard graph drawing tools~\cite{smoot_cytoscape_2011, bastian_gephi:_2009}.

\noindent \textbf{(3) Small multiples} show multiple instances of the same graph layout. Each instance encodes a different attribute dimension. Small multiples preserve the topology well, as they embed individual attributes directly in the graph~\cite{barsky_cerebral:_2008, lex_stratomex:_2012-1}. Disadvantages of small multiples include difficulty comparing attributes across the views, and having to render each individual graph with little space, limiting the size of the graph that can be visualized. 

\noindent \textbf{(4) Layout adaption} works by adjusting the layout so that a direct association between the nodes/edges and their attributes can be established. This is a broad category that includes placing the nodes in a scatterplot defined by two attributes as in GraphDice~\cite{bezerianos_graphdice:_2010} or aggregating nodes into bar charts as in GraphTrail~\cite{dunne_graphtrail:_2012}. Another strategy is to linearize (parts of) a node-link layout so that it can be easily juxtaposed with a table visualizing node or edge attributes. Examples of this approach include Pathline~\cite{meyer_pathline:_2010}, where a whole network including cycles and branching is linearized and juxtaposed with an attribute visualization; enRoute~\cite{partl_enroute:_2012}, which linearizes a user-chosen path; and Pathfinder~\cite{partl_pathfinder:_2016}, which queries for paths in networks and juxtaposes those paths with attribute visualizations. All these approaches make compromises between the readability of the topology of the graph and the association of the attributes to the network. 

\noindent \textbf{(5) Adjacency matrices} have both favorable and unfavorable properties compared to node-link layouts when judging topology~\cite{ghoniem_readability_2005}. Various attempts have been made to combine node-link layouts with matrices to find a compromise between these trade-offs. Examples are NodeTrix~\cite{henry_nodetrix:_2007}, which embeds adjacency matrices for subgraphs of a node-link layout, and MatLink~\cite{henry_matlink:_2007}, which enhances matrices with links. For attribute visualization, however, adjacency matrices are superior to node-link diagrams. For example, adjacency matrices can naturally encode edge attributes in matrix cells. Although this is mostly done with a single color value, multiple edge attributes can be visualized as nested graphs~\cite{elmqvist_zame:_2008}. Similar to the on-node encoding in node-link diagrams, however, the small space available for a matrix cell limits how much can be encoded. For node attributes, in contrast, it is easy to juxtapose multiple attribute visualizations with the rows or columns of the matrix. This has been done, for example in Graffinity~\cite{kerzner_graffinity:_2017} and in MapTrix~\cite{yang_many-to-many_2017}. 

Juniper is a layout adaption technique. It uses a linearized spanning tree to visualize a graph and juxtaposes it with a tabular visualization technique. We argue that this combination hits a sweet-spot in the topology-attribute trade-off spectrum. The linear tree-layout of the graph enables us to also juxtapose and align it with an adjacency matrix, resulting in a hybrid node-link/matrix technique, thereby leveraging the advantages of both: the ease of identifying paths in a node-link layout and the ability to quickly identify neighbors in the matrix layout.

\paragraph{Multivariate Tree Visualization}

Although the data we consider is of graph form, we present the graph as a tree. Hence it is useful to also consider the literature for multivariate trees in our review. Since trees are only a special type of graph, we can visualize it using any of these approaches.

In contrast to general graphs, trees can also be visualized using implicit layouts, such as tree maps~\cite{johnson_tree-maps:_1991}, sunburst plots~\cite{stasko_focus+context_2000}, or icicle plots~\cite{kruskal_icicle_1983}. Implicit techniques can use on-node encoding, such as color-coding on the node set, but they cannot be used to visualize edge attributes, as the edges are implicit.

A large number of techniques visualize attributes of the leaves of a tree in a tabular layout (a layout adaption strategy). Common examples are cases where the tree is a dendrogram that visualizes the hierarchical relationship of the items in a table~\cite{eisen_cluster_1998}. Similar approaches have been used for visualizing phylogenies and attributes about the species they contain~\cite{lee_phylodet:_2009, kreft_phyd3:_2017} or transactions associated with a hierarchy~\cite{burch_timeline_2008}. Surprisingly few techniques also visualize attributes for inner nodes in a tree. One example is a tree-table as it is used, e.g., in file browsers, showing properties such as file types and file/directory sizes. Another example is our Lineage tool~\cite{nobre_lineage:_2018}, which is designed to visualize clinical genealogies. The genealogies considered in Lineage are trees that are juxtaposed with a table that visualizes the properties of individuals. In some sense, Juniper is a generalization of the multivariate tree visualization techniques introduced in Lineage to general, highly connected graphs. Compared to Lineage, Juniper focuses on techniques that enable the exploration of a multivariate graph as a tree, which includes complete control over which edges to include in the tree, visualizing selected edges in an adjacency matrix, and dynamically growing the tree from a much larger graph. Section~\ref{sec:discussion} contains a detailed discussion of the differences of Juniper and Lineage.

\subsection{Tree-based Graph Visualization}
\label{sec:rw_tree}

The idea of tree-based graph drawing goes back at least two decades. Munzner uses a spanning tree as the structure to lay out a graph in hyperbolic space~\cite{munzner_drawing_1998} and shows links that are not part of the tree on demand. Hao et al.~\cite{hao_web-based_2000} take a similar approach, but they also introduce duplicates to resolve some ambiguities. Similarly, Ontorama~\cite{eklund_ontorama:_2002} uses a hyperbolic layout for a spanning tree and supplements it with a second view showing a linear tree that allows duplicate nodes. 

Yee et al.~\cite{yee_animated_2001} introduce a radial layout for graphs based on spanning trees. A focus node is used as the root of a spanning tree and shown at the center, immediate neighbors are shown circling the focus nodes, neighbors once removed are shown on a second circle, etc. The edges of the spanning tree and other non-tree edges are shown in a different color. Animated transitions are used to dynamically update the focus node. MoireGraphs~\cite{jankun-kelly_moiregraphs:_2003} follow the same principle but combine the radial layout with rich on-node attribute visualizations. 

The works most closely related to ours are TreePlus by Lee et al.~\cite{lee_treeplus:_2006-1} and the application-specific variant of TreePlus, GOTreePlus~\cite{lee_gotreeplus:_2008}. TreePlus introduces the ``plant a seed and watch it grow'' principle. Based on an initial, user-chosen node, analysts can grow the spanning tree by successively revealing subtrees. TreePlus shows hidden links between the tree nodes on demand using a combination of highlighting, a separate view of neighboring nodes, and explicit cross-links. Lee et al.\ evaluated TreePlus by comparing it to a traditional node-link diagram in a controlled study and found that TreePlus outperforms the node-link layout for most tasks and is preferred by most participants. For a detailed discussion of the differences of TreePlus and Juniper, refer to Section~\ref{sec:discussion}. Most of these techniques, including Munzner's hyperbolic tree, the radial layouts, and TreePlus, also encode node attributes, but they limit attribute visualization to on-node encoding of one or few attributes. 



Another type of technique visualizes compound graphs that have both a tree and a secondary graph structure. Fekete et al.~\cite{fekete_interactive_2003}, for example, visualize a tree structure in a compound graph as a tree map and render cross-links between the tree nodes on top of it. Holten~\cite{holten_hierarchical_2006} uses a compound graph as an example for his hierarchical edge bundling technique. Gou and Zhang~\cite{gou_treenetviz:_2011} render a tree structure in a sunburst layout and supplement edges connecting different levels of the layout. 

Although Juniper builds on this rich body of prior work, it is unique with regard to several aspects. Juniper leverages novel interactions and the close integration of tree-based graph visualization with an adjacency matrix to better support topology-based tasks in tree-based layouts. However, the main distinction of Juniper is the integration of an attribute table to support attribute-based tasks. The tree-based graph visualization techniques discussed here are limited to one or two attributes, in contrast to Juniper, which is the first tree-based graph visualization technique designed to handle highly multivariate graphs. 

\subsection{Query-based Visualization of Large Graphs}

A common strategy to explore large graphs is a bottom-up approach, where the analysis begins with a search or a query, and then more context is added as needed~\cite{vanham_search_2009, vonlandesberger_visual_2011}. Flavors of this approach range from explicitly revealing neighborhoods of nodes~\cite{heer_vizster:_2005, lee_treeplus:_2006-1}, to querying for paths or connectivity in a network~\cite{partl_pathfinder:_2016, kerzner_graffinity:_2017}, to querying based on a degree-of-interest function~\cite{vanham_search_2009}, to associative browsing and complex queries~\cite{kairam_refinery:_2015-1, tu_graphcharter:_2013}. All these examples are designed to return or expand a single subgraph, in contrast to techniques such as VIGOR~\cite{pienta_vigor:_2018} that are used to analyze (typically structural) queries that return many different subgraphs. 
Although we do not contribute novel concepts to graph querying methods, we make use of many of these approaches.

\section{Concept}
\label{sec:concept}

\begin{figure*}[t]
  \centering
  \vspace{-3mm}
  \parbox{.15\textwidth}{
    \vspace{-42.5mm}
    \subfigure[Source Graph.]{\includegraphics[height=2cm]{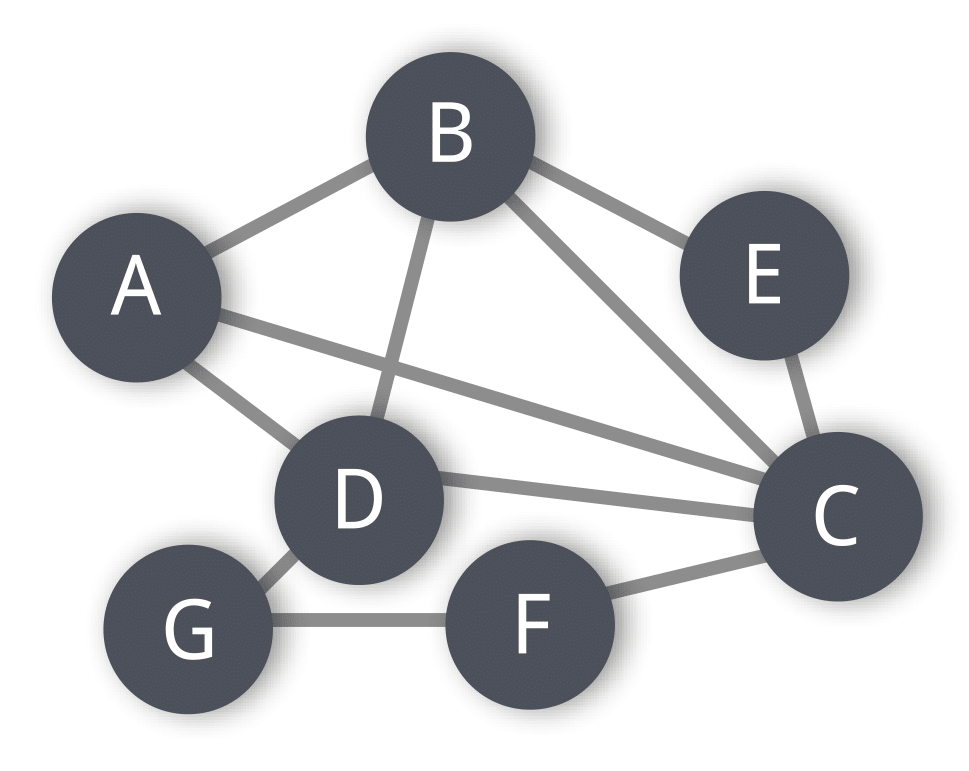} \label{fig:concept_graph} \vspace{-10mm}}\\ 
    \subfigure[Spanning Tree.]{\includegraphics[height=2cm]{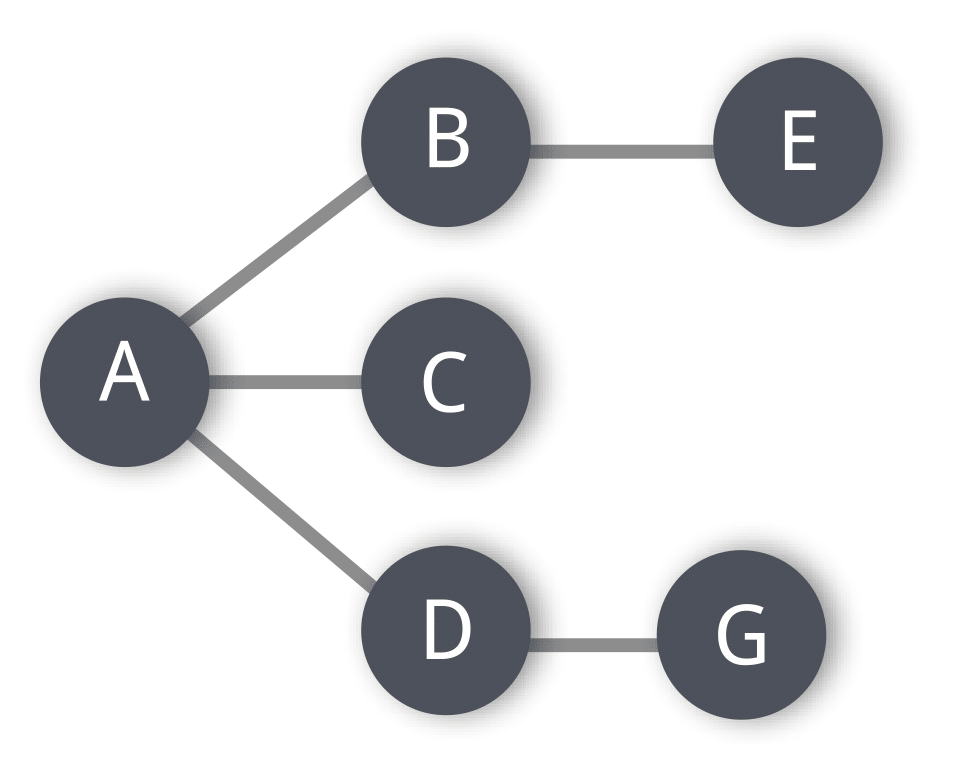} \label{fig:concept_tree}}
  } 
  \hfill
  \subfigure[Linearized Tree Layout with Tables.]{\includegraphics[height=5cm]{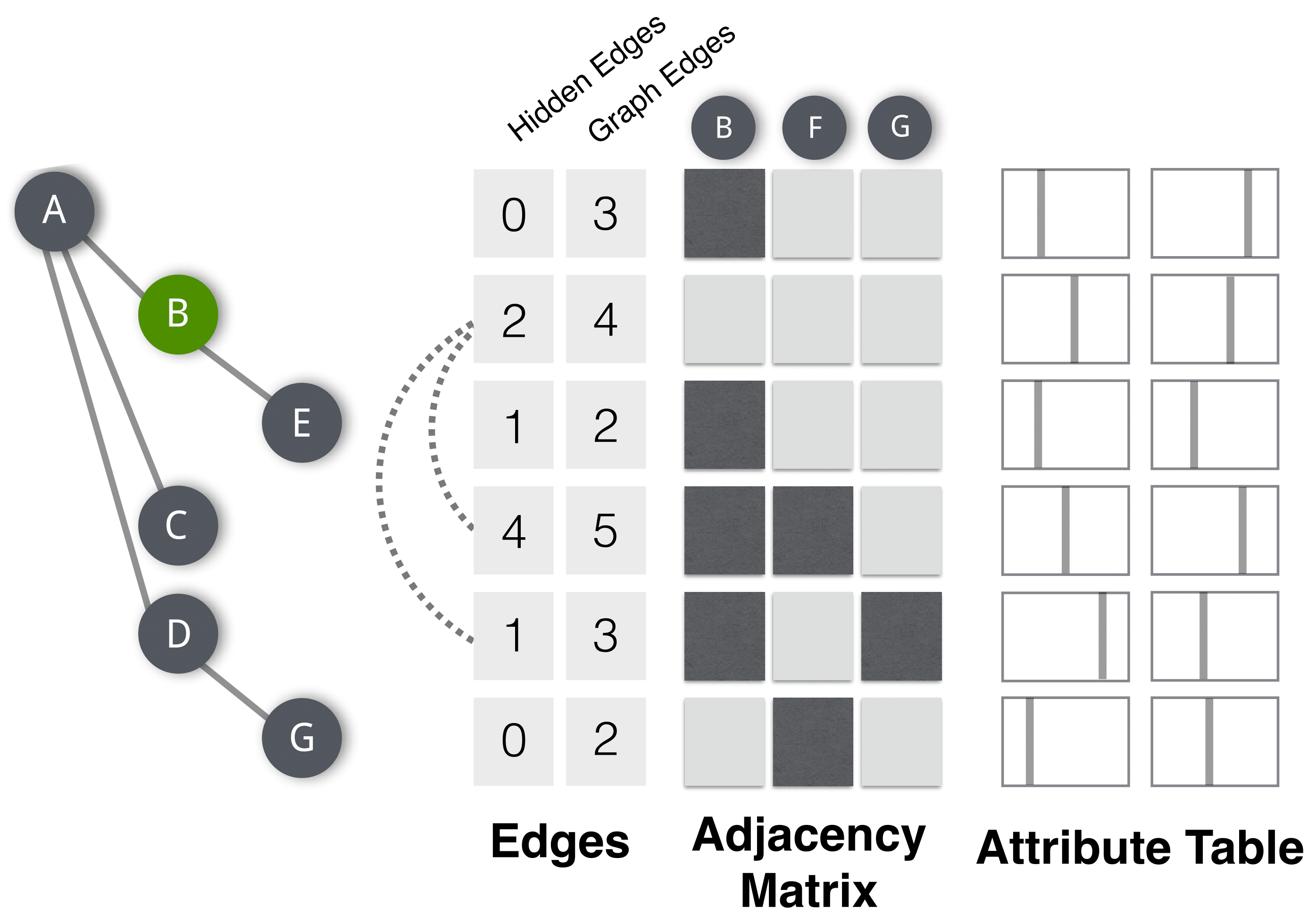} \label{fig:concept_linearization}}
  \hfill
  \subfigure[Level Layout.]{\includegraphics[height=5cm]{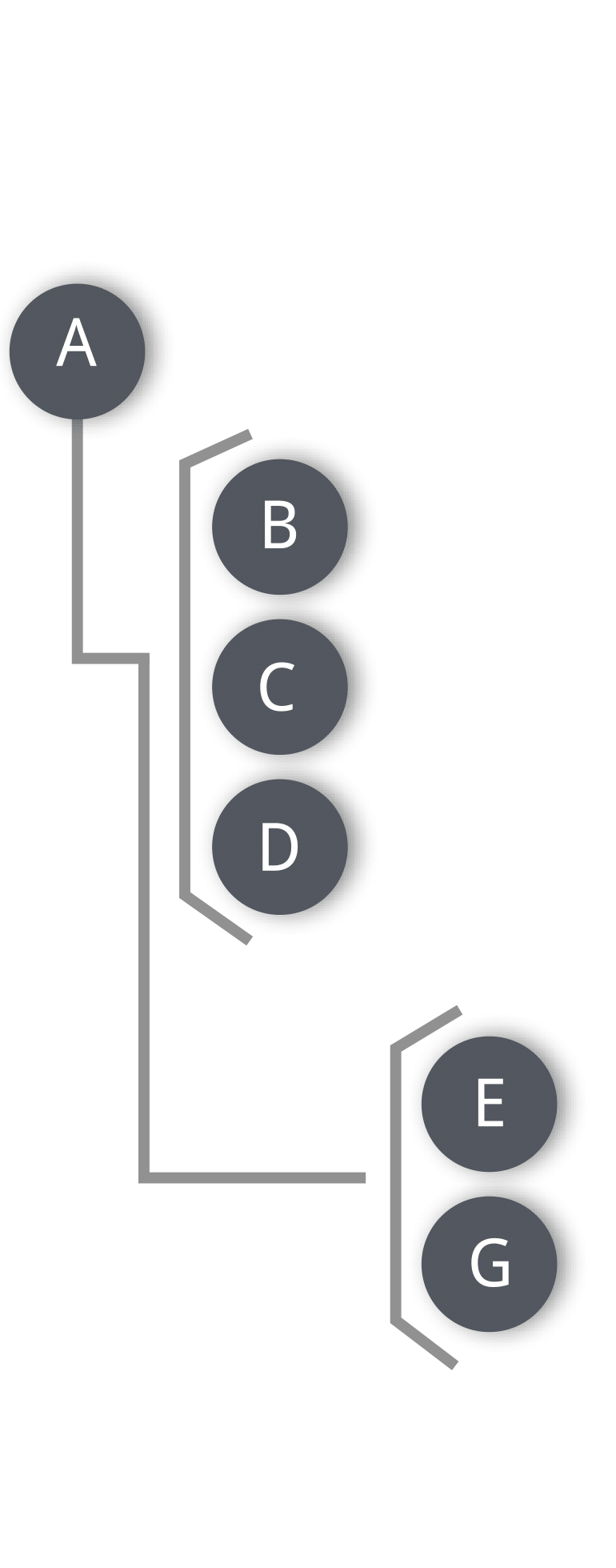}\label{fig:concept_level}}
  \hfill
  \subfigure[Path-Based Sorting.]{\includegraphics[height=5cm]{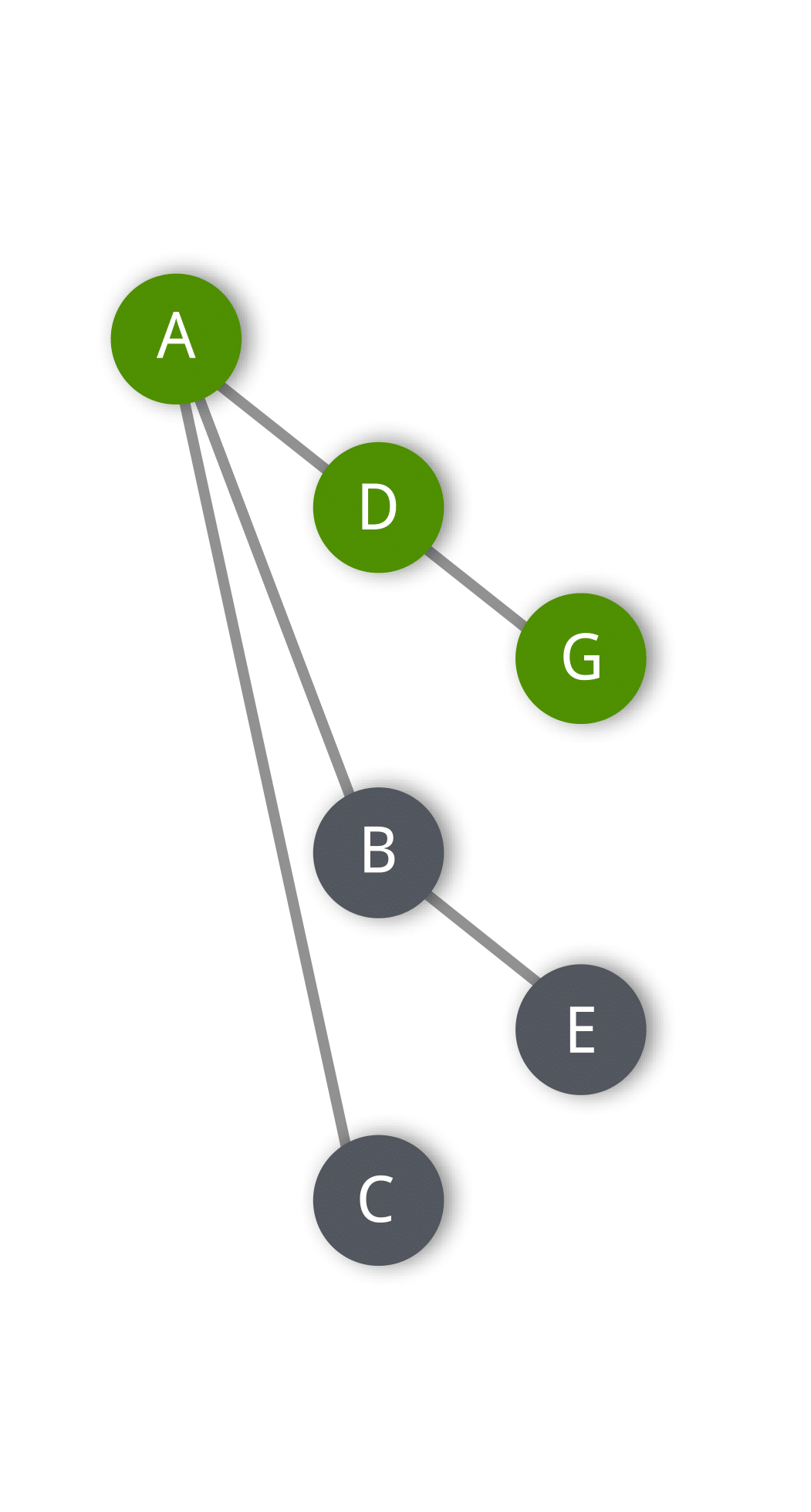} \label{fig:concept_node_sorting}}
\vspace{-3mm}
  \caption{From (a) a graph, to (b) a spanning tree of a subgraph. Note that node F is not included and that several edges are missing (e.g., B-D). (c) Linearization of the tree shown in (b). The linear tree layout allows us to juxtapose a table showing hidden edges and overall node degree, an adjacency matrix, and a table showing rich node attributes. Hidden links are shown for the selected node B. (d) Level layout of the same tree, where all nodes at the same distance from the root are grouped together. (e) Node sorting to ensure that all nodes on path A-D-G are in sequence.}
   \label{fig:concept_graph_to_tree}
  \vspace{-5mm}
\end{figure*}

In this section we introduce the concept of tree-based exploration of multivariate graphs. Details on our implementation of this concept and a number of design decisions can be found in Section~\ref{sec:design}. 

The idea that we follow is to (1) extract a subgraph from a larger, underlying graph, (2) calculate a spanning tree from the subgraph, and (3) linearize this tree. The linearization enables us to juxtapose the tree with a table, as illustrated in Figure~\ref{fig:concept_graph_to_tree}. This tree+table approach, in turn, allows us to visualize additional topological information, such as node adjacency, and to show associated attributes of the nodes. Although the first two steps are common in other systems, as discussed in Section~\ref{sec:rw_tree}, Juniper is the first technique to make use of a dynamically extracted tree to visualize multivariate attributes. 



Figure~\ref{fig:concept_graph} shows an example graph. In practice, this graph can be larger than can be conveniently displayed, can have different types of nodes, and can have rich attributes associated with it. Following the ``search, show context, expand on demand'' principle~\cite{vanham_search_2009}, we extract a subgraph from the larger graph --- either in bulk or iteratively --- and calculate a spanning tree for that subgraph using a breadth-first search (Figure~\ref{fig:concept_tree}). 
If a subgraph is added in bulk, a key decision in this process is the choice of the root node, since the tree-based approach works best for tasks related to the root (e.g., it is trivial to see all neighbors of the root). We assume that analysts will want to manually specify a root in most cases; if no root is specified, we choose the node with the highest degree. The order in which nodes are visited at a given level by the breadth-first search algorithm also has an impact on the resulting tree, as nodes visited first will likely have more of their neighbors available to be attached. In Juniper, the order is driven by a user-defined sorting function; sensible options include lexical ordering of node labels, ordering by degree, or ordering by attributes. 

\subsection{Layout}
Once a spanning tree is calculated, we linearize the tree using one of two complementary layout algorithms. We produce a traditional \textbf{tree layout} using a depth-first search algorithm, where every node is assigned a unique vertical position (see Figure~\ref{fig:concept_linearization}). The order of nodes for layout purposes is again defined with a sorting function.

An alternative layout is the \textbf{level layout}, shown in Figure~\ref{fig:concept_level}. In the level layout, all nodes of a level are shown next to each other, followed by all nodes of the next level, etc. Again, sorting of nodes is driven by a user-specified function. 

Level layout and tree layout have complementary strengths. The tree layout is well suited to investigate precise relationships to the root node. For example, in the bipartite co-author network, if we start with an author, we can expand all her publications, and then expand all the co-authors on each of these publications, giving us a sense of who collaborated on which paper. The level layout, in contrast, allows us to ask a different question. In the level layout, the root author would be at level one, all her papers at level two, and all her co-authors at level three. In this layout, we can easily see and compare all the co-authors of the root author; they will be next to each other, and we can use the table to sort the nodes, to identify, for example, the author with the most papers. In general, the tree layout can be used to answer questions about specific topology, whereas the level layout can be used to evaluate all nodes at a certain distance. Note that level and tree layouts can be separately defined for each branch. 

Both level and tree layouts are well suited to support one of our main tasks: understanding attributes in the context of neighborhoods. To support our other main task --- understanding attributes in the context of paths --- we introduce path-based node sorting as illustrated in Figure~\ref{fig:concept_node_sorting}. In this example, the tree shown in Figure~\ref{fig:concept_tree} was reordered to guarantee that all nodes along the path A-D-G are in the sequence of the path, thereby supporting the analysis of its attribute in sequence and enabling analysts to make judgments about path effects. 

\subsection{Reshaping the Tree and Revealing Hidden Edges}

The crossing-free and easily readable layout achieved by using a spanning tree comes at a cost --- both tree and level layouts hide edges. A simple way to reveal all edges and neighbors of a node is to make it the root. This, however, changes the layout drastically, which can be disorienting for an analyst. An alternative is to \textit{gather all children} of a node. In that case, all nodes that have an edge to the target node are attached as children to this node, with the exception of its ancestors. We choose not to attach ancestors as children because it would lead to similar layout changes as the make-root operation. 

In addition to reshaping, we use three strategies to visualize edges that are not part of the tree. First, hidden edges are drawn for user-selected nodes. In Figure~\ref{fig:concept_linearization}, hidden edges are drawn for node B, which has edges to nodes C and D, in addition to the edges to A and E that are part of the tree. This strategy is common to most tree-based graph visualization techniques (e.g., \cite{lee_treeplus:_2006-1}). 

Complementary to showing hidden edges on demand, we also show a table visualizing counts for hidden edges (the number of hidden edges in the subgraph) and graph edges (the degree of the node in the underlying graph), as shown in Figure~\ref{fig:concept_linearization}. 
Whereas the former allows analysts to judge connections that are not apparent in the tree, the latter can be used to judge the node relative to the whole network, and also give analysts a sense of how many nodes would be added if the neighbors of the node were to be added to the subgraph. 

The third strategy to visualize topology is an adjacency matrix that is fully integrated with the tree, resulting in a hybrid node-link/matrix layout. The matrix is not meant to show all nodes in the subgraph, as this would likely result in a sparse matrix and require considerable amounts of screen-space. Instead, similar to the rationale behind NodeTrix~\cite{henry_nodetrix:_2007}, the matrix is designed to show connectivity for highly connected nodes. The integration of the node-link tree and the matrix allows analysts to quickly judge the relationships of these nodes with nodes in the tree. Note that any node can be included in the adjacency matrix, not only those that are part of the subgraph. Figure~\ref{fig:concept_linearization}, for example, shows node F in the adjacency matrix, which is not included in the subgraph. The adjacency matrix can be useful, for example, when exploring an author's papers and co-authors. Adding the author's PhD and postdoc advisors to the matrix is useful since she has likely collaborated on many papers. Using the matrix, an analyst can quickly judge which papers were written in collaboration with whom, which would not be easy to see in the tree visualization alone. 

\subsection{Hiding and Aggregation}
\label{sec:aggregation}


\begin{figure*}[t]
  \centering
  \vspace{-3mm}
    \hspace{1cm}
    \subfigure[Aggregating a Tree Layout.]{\includegraphics[height=4.5cm]{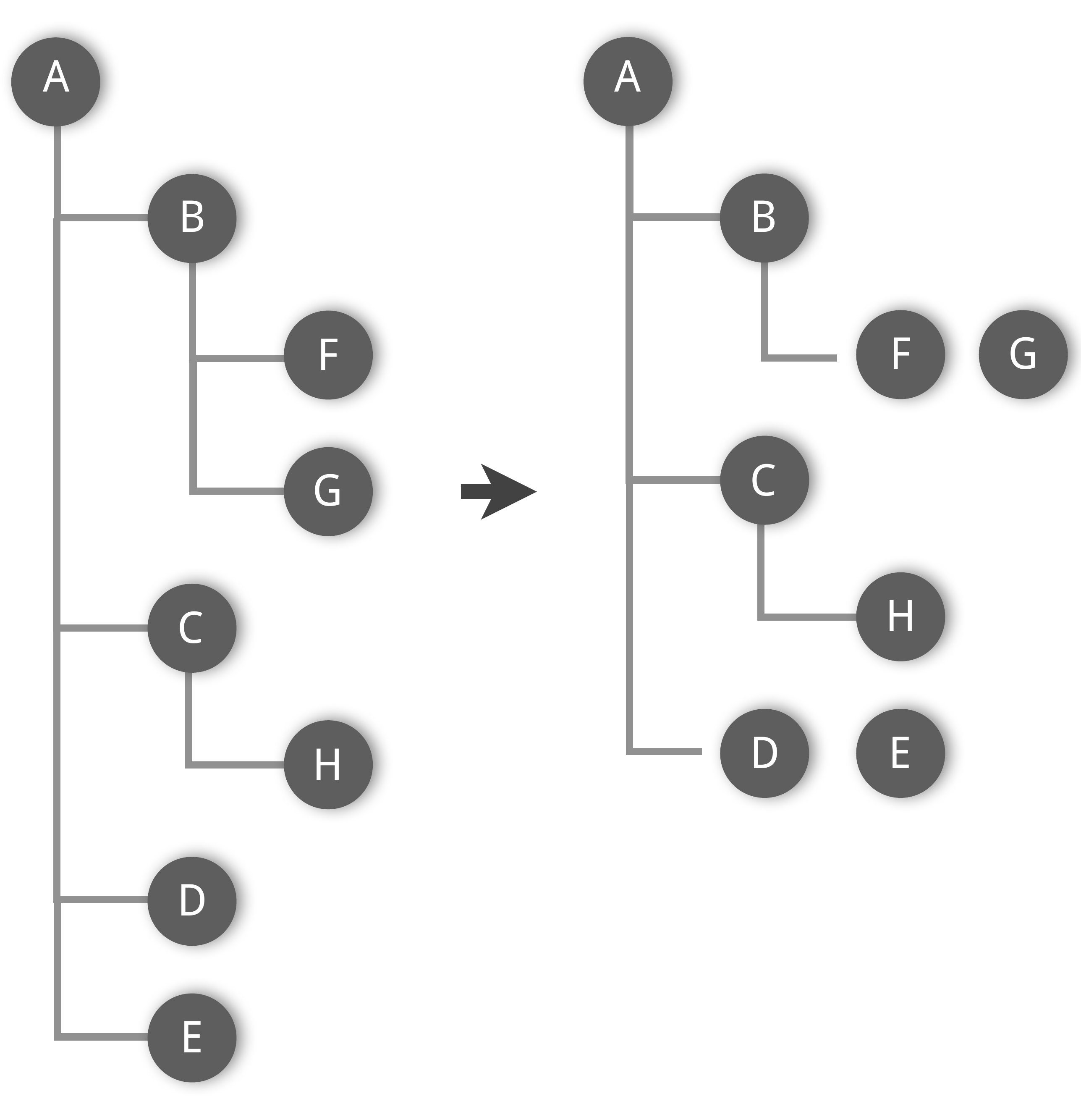}
    \label{fig:agg_tree}}
    \hspace{2cm}
    \subfigure[Aggregating a Level Layout.]{\includegraphics[height=4.5cm]{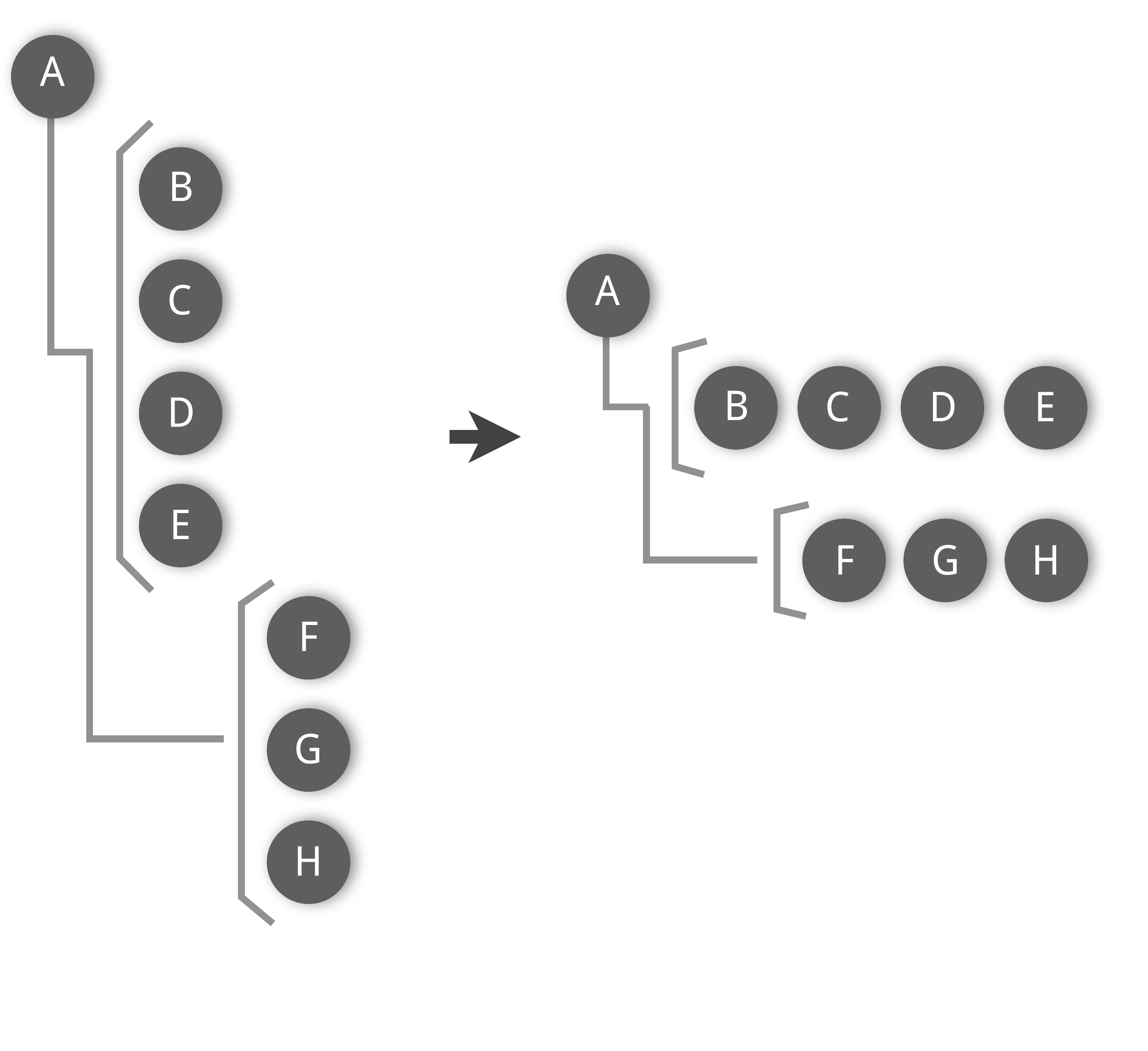} \label{fig:agg_level}}
    \hfill
    \subfigure[Aggregating with DOI.]{\includegraphics[height=4.5cm]{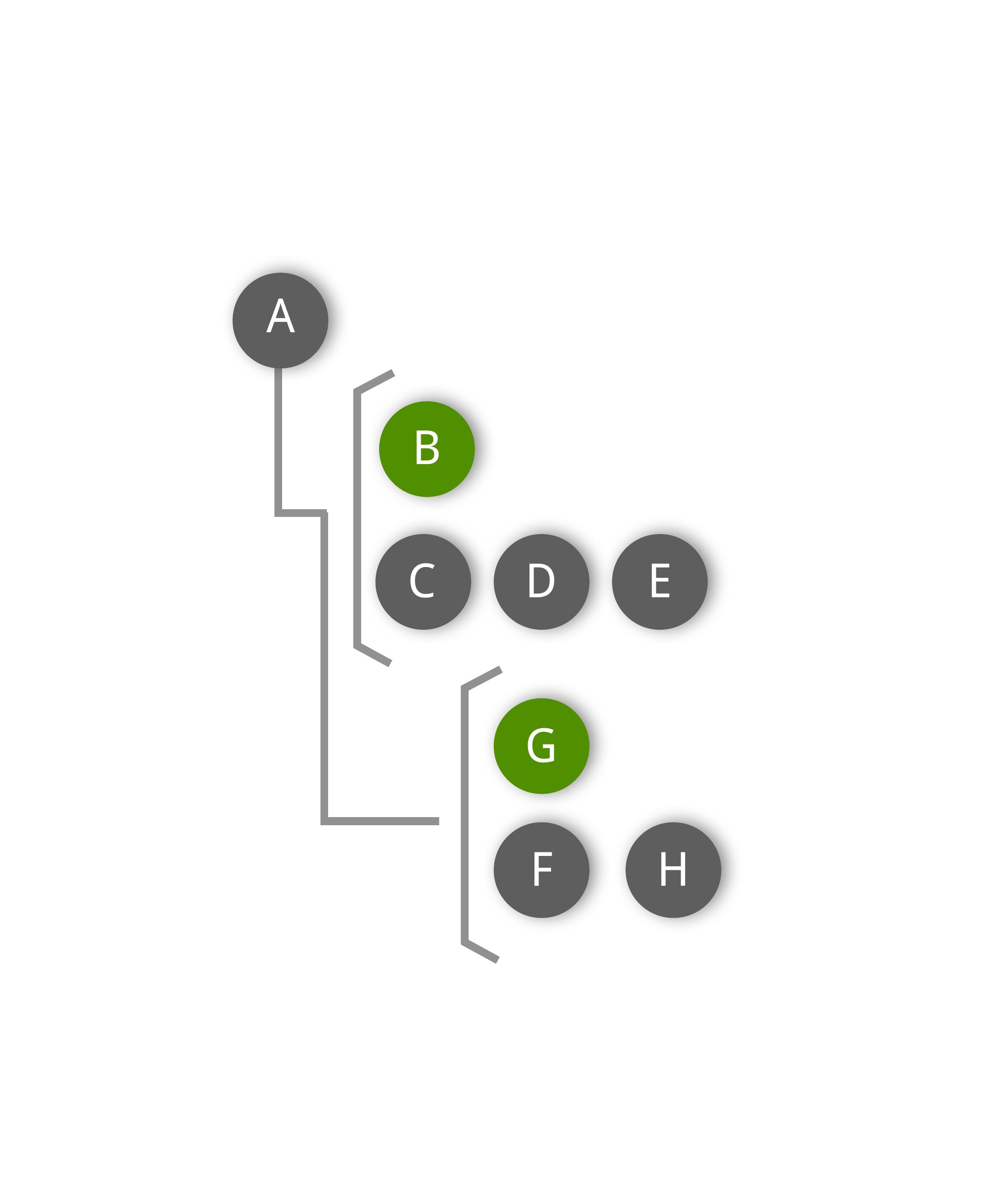} 
    \label{fig:agg_doi}}
        \hspace{5mm}
  \vspace{-3mm}
  \caption{Aggregation strategies. (a) Aggregation in tree layout: leaves of the same parent are aggregated by placing them in the same row. (b) Aggregating in level mode: nodes of the same level are aggregated into a single row. (c) Aggregation with a degree-of-interest function, shown in a level mode. Nodes B and G (green) are considered to be of interest based on a degree-of-interest function, and hence are placed in their own row.}
   \label{fig:agg}
  \vspace{-5mm}
\end{figure*}

Although a tree-based linear layout has many advantages, it also limits the number of nodes that can be concurrently displayed on the screen. To counteract this limitation, we introduce two approaches to selectively reduce the number of nodes: branch hiding and branch aggregation.  

Branch hiding is common to most tree visualizations. It allows analysts to selectively hide branches of a tree that may not be relevant for the task at hand. Although it excels at saving space, the downside of hiding is that analysts no longer have access to any information about the hidden nodes. 
    
Our second, less aggressive, approach is aggregation. This approach has the advantage of preserving both topological and attribute information in aggregate form. Aggregation is available in both tree and level modes. In tree mode, illustrated in Figure~\ref{fig:agg_tree}, only leaves are aggregated; the backbone of the tree and hence all the topological structure of the tree remain visible. When aggregating in level mode, as shown in Figure~\ref{fig:agg_level}, all the nodes of one level are aggregated into a single row, resulting in a very compact layout. 

Aggregation as described above can be controlled using the tree's topology, i.e., analysts can choose to represent individual branches in aggregated mode. However, it is a common task to look for nodes with certain attribute characteristics among such a large, aggregated set. To address this, we introduce a binary degree-of-interest (DOI) function~\cite{furnas_generalized_1986}. Figure~\ref{fig:agg_doi} illustrates the effect of a DOI function on level-based aggregation. Here two nodes, B and G, shown in color, are considered of interest and hence retain their own row, whereas the others are aggregated. An example for the co-author network would be to look for all highly cited papers of a network of prolific authors, in which case highly cited papers would be afforded their own rows, whereas papers with few citations would be aggregated.  

Note that the visualization of hidden edges, the adjacency matrix, and the attribute visualization can easily be adapted to support both individual and aggregated rows. 
    
\subsection{Attribute Visualization}

In line with the ``topological attributes'' shown in the edge count table and the adjacency matrix, we can leverage the linearized layout to visualize arbitrary node attributes, as illustrated in Figure~\ref{fig:concept_linearization}. A variety of visualization options can conceivably show different data types in either individual or aggregated form~\cite{furmanova_taggle:_2018}.

The key benefit of the integrated attribute visualization, as opposed to a separate linked view, is that the topology of the tree can be used to sort and group the elements, revealing, e.g., dependencies along a path, or shared characteristics of all neighbors of a node. Equally valuable is the opposite approach: the attribute visualizations can be used to influence the tree layout, through both sorting and DOI functions. The columns representing attributes are well suited to interactively define such a sorting, or a data range of interest for a DOI, as shown in Figure~\ref{fig:design_demo}. 
    
\section{Design}
\label{sec:design}

We implemented the concept described in the previous section in an interactive web-based tool. Here we report on the design decisions that went into realizing this tool. 

Juniper has two views: (1) a query view that is used to search for individual nodes or to query for subgraphs, and (2) the main tree+table view, which contains the graph and attribute visualizations. A toolbar at the top allows analysts to switch to a force-directed layout and to choose a dataset. In addition, node-type specific menus allow analysts to add attributes to the table and to filter nodes by type. 

\subsection {Querying}

The query interface is the starting point for any exploration in Juniper.
Analysts can browse or search for nodes in the query view and add them to the tree+table view (see Figure~\ref{fig:juniper_overview}). The search and browsing interface is faceted by node types: when, for example, text is entered in the search field, all matches are shown in separate, type-specific facets. The faceting enables analysts to quickly find nodes of interest, even with an incomplete query. The interface also shows the degree of the nodes, so that highly connected nodes can be readily identified. A node can be added individually or together with all its neighbors. 
Nodes can also be added to the adjacency matrix. It is possible to add multiple roots/trees to the tree+table view simultaneously.


The query view also provides an interface to write Neo4J Cypher queries (a query language for the graph database we use). Although this is an expert option, it enables analysts to retrieve arbitrary subgraphs considering both topological features and attributes.  
    
\subsection {Tree View}

The tree view implements the concept outlined in Section~\ref{sec:concept}. Nodes at each level are given ample space for labels, which is a common limitation in force-directed layouts. We also distinguish between different node types by showing a custom symbol for each type. Edge types and directions are shown as tool-tips where available.

The graph can be grown organically by revealing neighboring nodes.
In cases where a node has more neighbors than are currently shown, a small plus sign is shown below the node that can be used to add those missing nodes, as shown in Figure~\ref{fig:design_demo}.

In terms of tree-restructuring, our prototype supports the previously discussed \textit{make root} and \textit{gather children} operations, in addition to selectively removing nodes/branches, and explicitly reattaching a branch at a different node, based on a hidden edge. Hidden edges are shown for the selected node, \house~\textit{House Stark}, in Figure~\ref{fig:design_demo}.


\subsubsection*{Layout and Aggregation}

\begin{figure*}[ht]
\centering
\includegraphics[width=\linewidth]{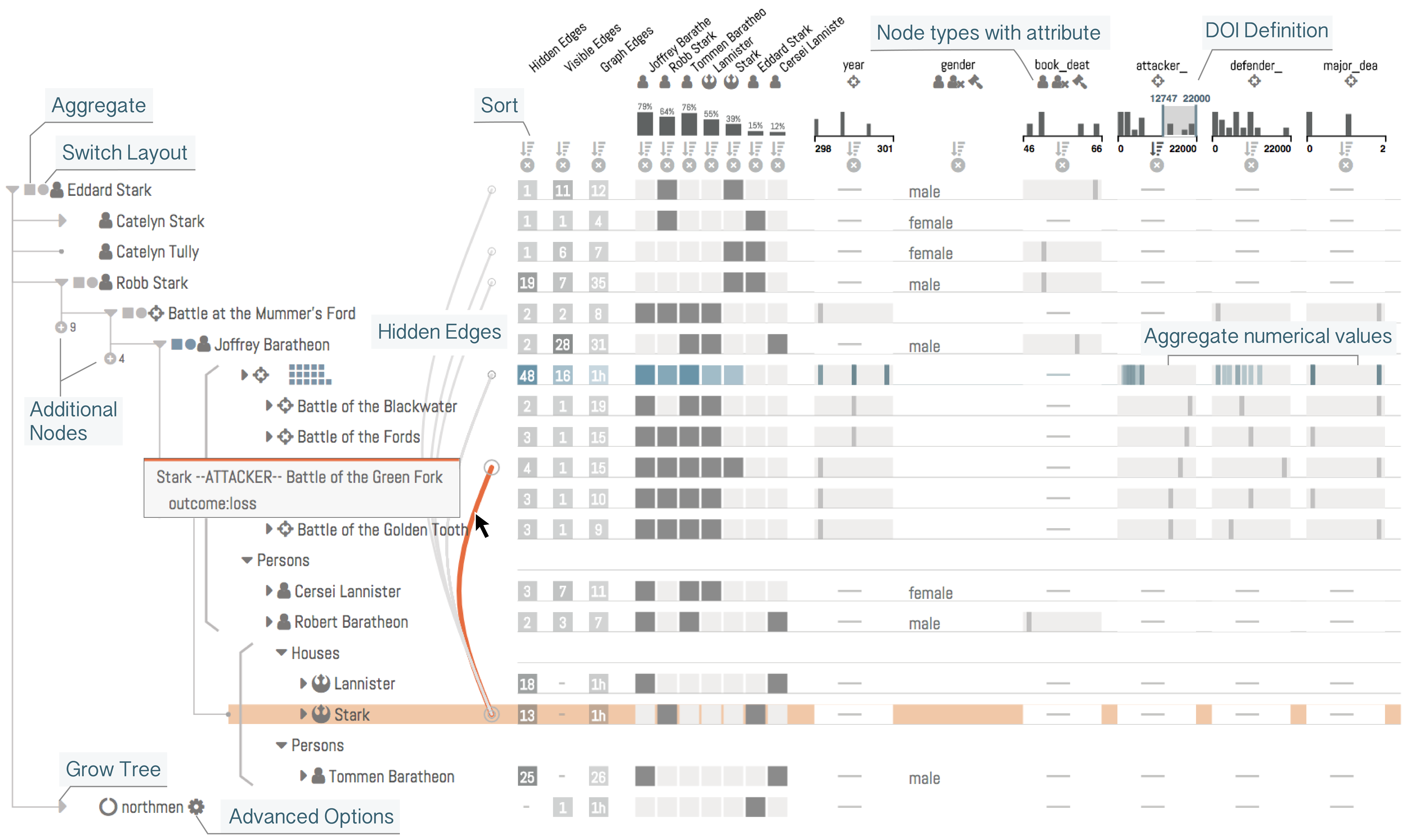}
\vspace{-7mm}
\caption{Juniper design overview using a Game of Thrones dataset, rooted at \textit{Eddard Stark}, and expanded in tree layout up to \textit{Joffrey Baratheon}. Descendants of the node Joffrey Baratheon are shown in level mode. A DOI function reveals \battle~battles with an attacker size of 10,000 and larger in the otherwise aggregated set of battles. The associated table visualizes edge counts (hidden, visible, and graph edges) for both individual nodes (gray) and aggregates (blue). The adjacency matrix was auto-populated with the most connected nodes in the subgraph. Again, individual rows are shown in gray, aggregates in shades of blue. The attribute columns are specific to node types, as shown in the column header. Aggregated rows use compact visualizations showing the values of all contained rows, where appropriate. Hidden edges are shown for the selected \house~\textit{House Stark}. The edge from House Stark to the \textit{Battle of the Green Fork} is highlighted and a tooltip with information about the edge type and direction is shown.}
\label{fig:design_demo}
\vspace{-5mm}
\end{figure*}

Figure~\ref{fig:design_demo} shows the implementation of the previously described layout strategies. The example shown is a Game of Thrones network that contains many different node types. The tree originates at a person, \textit{Eddard Stark}, who has a connection to the \battle~\textit{Battle at the Mummer's Ford} via an intermediate person, \textit{Robb Stark}. The layout at the root is a tree layout, but descendants of \textit{Joffrey Baratheon} are shown in level layout, which is indicated by the brackets replacing direct connections. 
Note that the branch starting at Joffrey is aggregated. We show aggregated nodes as little squares, which allows easy size estimation, and facet them by node type, so that analysts can quickly see how many nodes of a certain type are in each aggregate. Tooltips on the aggregated nodes reveal the node title. \person~Persons and \house~Houses are manually deaggregated; for \battle~Battles we use a degree-of-interest function to partially deaggregate battles above a certain \textit{attacker size} (see label ``DOI definition'' in Figure~\ref{fig:design_demo}).

\subsubsection*{Edge Count Table and Adjacency Matrix}

The table for edge counts described in Section~\ref{sec:concept} is realized with a redundant encoding using color saturation and exact numbers, as shown in Figure~\ref{fig:design_demo}. Numbers with three or more digits are shortened to the most significant digit plus `h' for hundred and `k' for thousand. Since aggregate values and individual nodes are commonly of different scales, we use separate color scales for individual rows (gray) and aggregates (blue). The color scales are defined independently for each column, since the number of hidden edges is expected to be much smaller than the number of graph edges, for example.

The purpose of the adjacency matrix is to further expose the connections in the graph that are not captured in the tree. As discussed in Section~\ref{sec:concept}, we do not show all nodes in the matrix column, but rather selected nodes that complement the tree well in a hybrid node-link/matrix layout~\cite{henry_nodetrix:_2007}. Nodes can be added to the table from the query view or the tree. We also auto-populate the matrix with the most connected nodes in the tree since highly connected nodes are likely to have many hidden edges. As in the edge count table, we use grayscale (binary in this case) for individual rows and a blue color scale for aggregate rows. In contrast to the edge count table, the color scales are normalized on a per-row basis to account for aggregates of different sizes.
    
\subsubsection*{Attribute Table}
        
The attribute table can be used to visualize rich data associated with the nodes. Each column in the table corresponds to an attribute for one or multiple node types. Most attributes are defined for only one node type, which can result in a sparse table if a graph contains many different node types. For numerical data we use a vertical line placed along a scale, as it uses position, the most powerful visual channel available. Exact values are shown on hover. We visualize aggregate rows by drawing multiple lines in the same cell, as shown for the aggregate cell for \battle~battles and defender size column in Figure~\ref{fig:design_demo}, for example. By using transparency, we can ensure that overlapping lines are noticeable. 

The attribute table, the edge count table, and the adjacency matrix also serve as interfaces for sorting and defining degrees of interest, as shown in Figure~\ref{fig:design_demo}. Sorting is applied only within the levels of the tree to avoid edge crossings. Columns can be arranged arbitrarily through drag and drop.
    
\subsubsection*{Path Visualization}

\begin{figure}[t]
  \centering
  \vspace{-1mm}
    \subfigure[Path Preview]{\includegraphics[width=\linewidth]{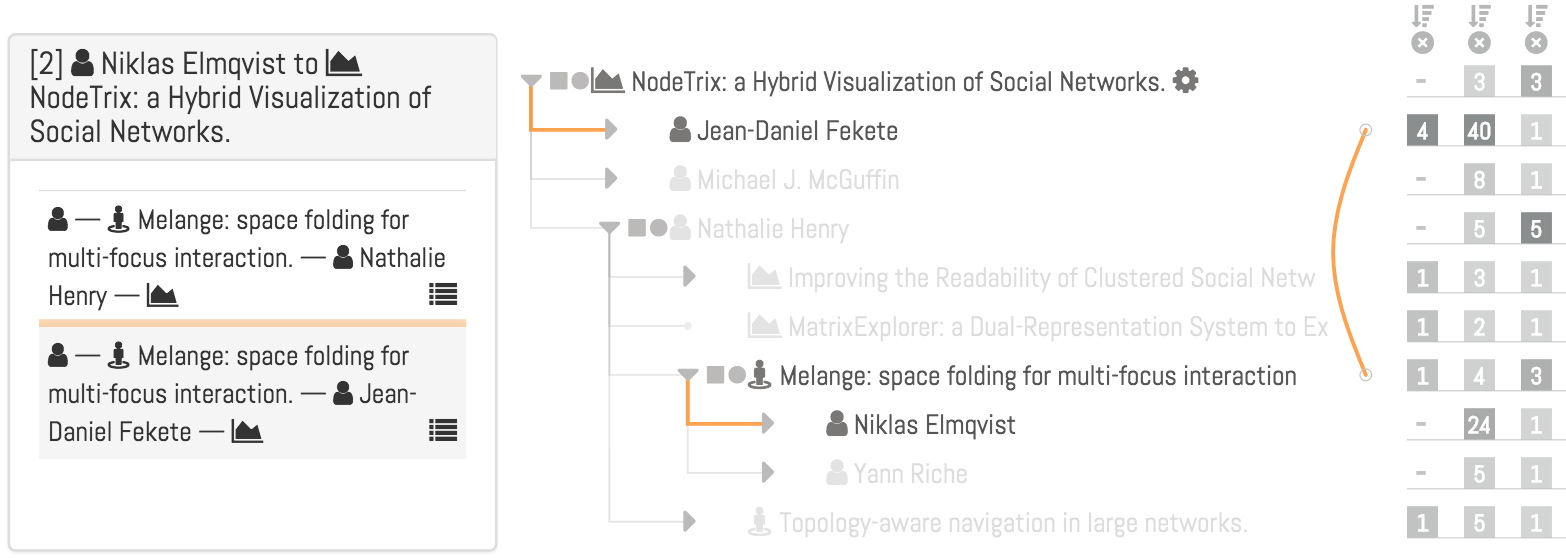}
    \label{fig:path_preview}}\\ \vspace{-2mm}
     \subfigure[Sequential Path]{\includegraphics[width=\linewidth]{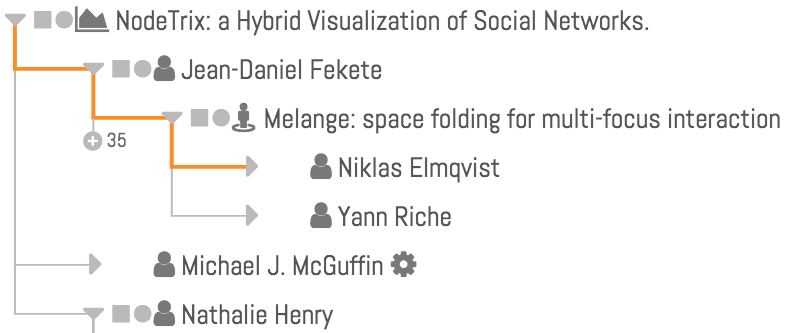} 
    \label{fig:path_sequential}}
  \vspace{-3mm}
  \caption{Shortest path search and visualization. (a) A shortest path search between \textit{Niklas Elmqvist} and the \textit{NodeTrix} paper reveals two paths, shown in the view on the left. Both paths go through the \textit{Melange} paper, but one continues through \textit{Nathalie Henry}, the other through \textit{Jean-Daniel Fekete}. The path via Jean-Daniel Fekete uses a hidden edge, which is shown when hovering over the path. (b) The selected path was laid out sequentially.}
   \label{fig:path}
  \vspace{-5mm}
\end{figure}

A common task in networks is to find a short(est) path between two nodes~\cite{partl_pathfinder:_2016}. Since shortest paths can be hidden when using a tree-based layout, Juniper provides an explicit path search feature to quickly identify all shortest paths between two nodes (Figure~\ref{fig:path}). A dedicated view lists all the paths of the same length. Note that this list is limited to paths in the subgraph, but it could easily be extended to the whole graph. When hovering over a path, the path is highlighted in the tree and shows hidden edges if necessary. On demand, analysts can enforce that all nodes in a selected path are laid out sequentially in the tree and, by extension, in the table (Figure~\ref{fig:path_sequential}). This is an example of how topological features can be used to lay out the attribute table to study potential network effects in attribute space. 

\section{Implementation}

Our prototype can be accessed at \url{http://juniper.sci.utah.edu/}. Juniper is implemented as a web-application using Typescript and D3 on the client and Python and Flask on the server. Deployment and plugins are managed using the Phovea framework\footnote{\url{https://github.com/phovea/}}. The graph data is stored in multiple Neo4J\footnote{\url{https://neo4j.com/}} graph databases, each running in a separate Docker container on designated ports. Graph queries can be submitted directly through the advanced query interface or they are more commonly exposed through a REST API.

Juniper is open source and uses the permissive BSD license. The source code is available at \url{https://github.com/caleydo/lineage/tree/juniper}. 

\section{Examples and Use Cases}
Here we show several explorations using Juniper for focus tasks. We demonstrate how both attribute and network data are used in conjunction to gain insights. 

\subsection{Game of Thrones Network}

The network we use in this example is based on the popular books and television show ``Game of Thrones'' or ``A Song of Ice and Fire'' by George R. R. Martin. The dataset is available on Kaggle\footnote{\url{https://www.kaggle.com/mylesoneill/game-of-thrones/}}. We followed instructions to import them into Neo4j\footnote{\url{https://tbgraph.wordpress.com/2017/06/25/neo4j-game-of-thrones-part-3/}}. The network captures several types of relationships among story characters, noble houses, battles, books, cultures, etc. The network contains about 2,500 nodes, 17,000 edges, 18 attributes, and 11 node types.


We start our exploration with one of the main characters in the show, \textit{Eddard Stark}. We see that he is associated with a handful of people, as well as all five of the books. The books are a hub of connectivity with ties to most characters in the story. As this is not helpful for our investigation, we filter out nodes of type book. We know that Eddard was killed by \textit{Joffrey Baratheon}, so we add him to the tree through the search interface. Surprisingly, the dataset does not capture this direct connection between Eddard and Joffrey. We see that they are connected through a set node (both are nobles) instead, which is not interesting, so we filter out these nodes, leaving us with a connection between the Starks and Joffrey through the \textit{Battle of the Mummer's Ford}, as shown in Figure~\ref{fig:design_demo}. Next, we are interested in seeing all of Joffrey's connections, so we use the gather children operation. We see that he is connected to several battles, a few people, and to \textit{House Lannister}. 

We want to get a better understanding of the battles and what the role of the opposing houses of Stark and Lannister is in them, so we switch the branch starting at Joffrey into level mode, which groups all of his node's descendants by their type. We then aggregate battles, add several attributes related to battles to the table and inspect them. We see in the \textit{attacker size} aggregate cell that battles seemed to fall into one of two groups: a few large battles with an attacker size of over 10,000 people and many smaller battles. We are particularly interested in understanding the large battles, so we use a brush on the histogram for that column to set a DOI function to reveal all battles with an attacking force of more than 10,000 people, as shown in Figure~\ref{fig:design_demo}.  

Because we are interested in the involvement of the Stark and Lannister houses in these large battles, we hover over each independently to see their connections to the battles. We see that the Stark house is associated with only one of these large battles --- the \textit{Battle of the Green Fork}. Hovering over this edge reveals that House Stark was the attacker and lost this battle. Inspecting the adjacency matrix cell that connects House Stark to the aggregated smaller battles shows us that House Stark was associated with nine of the 16 battles with an attacker size of under 10,000. Clearly, the direct interaction between Lannisters and Starks in battle happened mainly in smaller battles. 

\subsection{Exploring a Co-Author Network}


\begin{figure*}[t]
  \centering
  \vspace{-3 mm}
    \subfigure[Distribution of papers by authors of TreePlus.]{\includegraphics[width=0.58\linewidth]{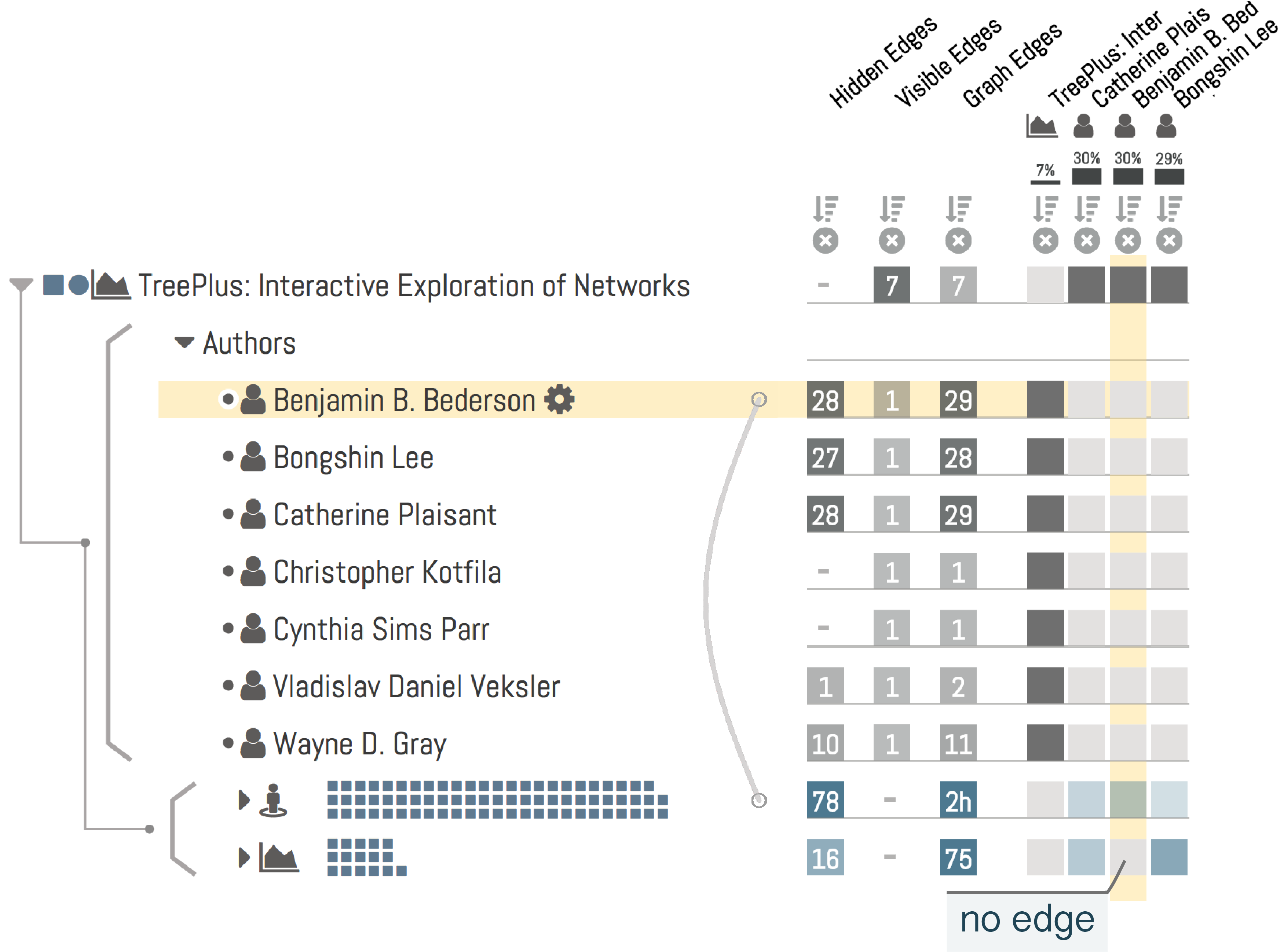}
    \label{fig:co-auth:grouped_papers}}
    \subfigure[Individual paper distributions.]{\includegraphics[width=0.35\linewidth]{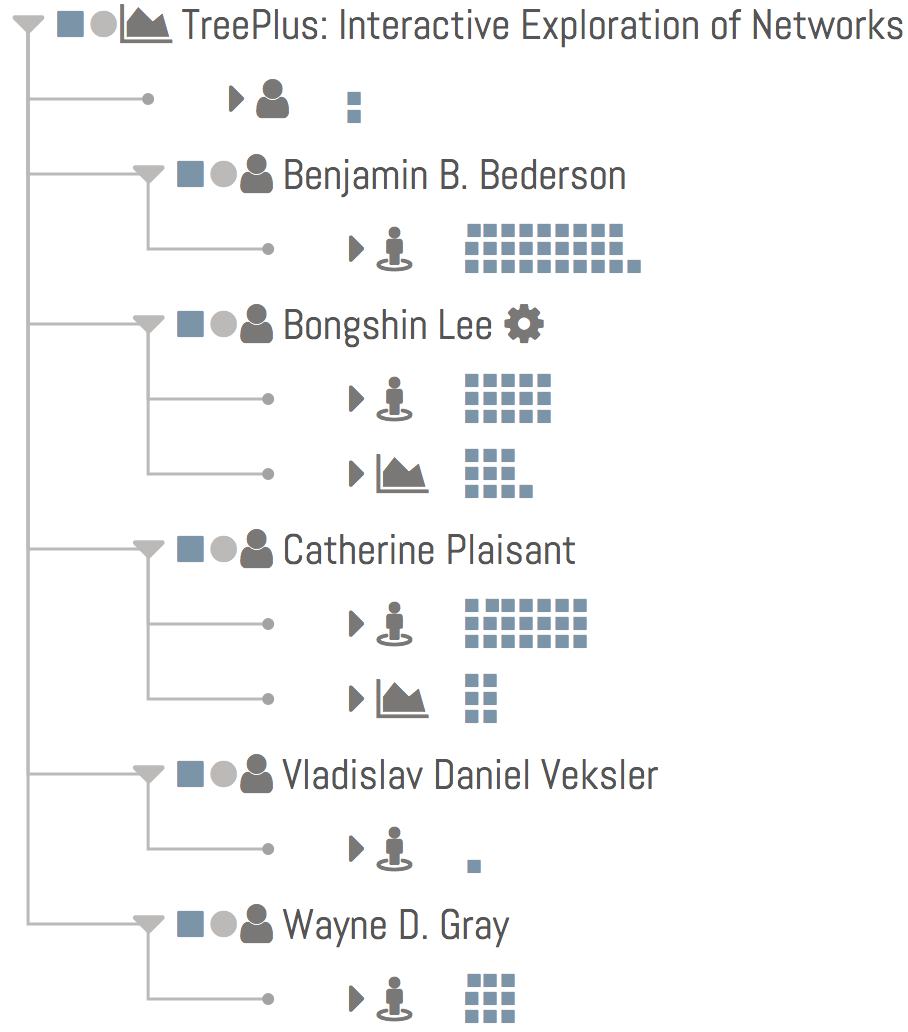} \label{fig:co-auth:author_conf_dist}} \\ \vspace{-3mm}
    \subfigure[Frequent co-authors of Catherine Plaisant.]{\includegraphics[width=0.95\linewidth]{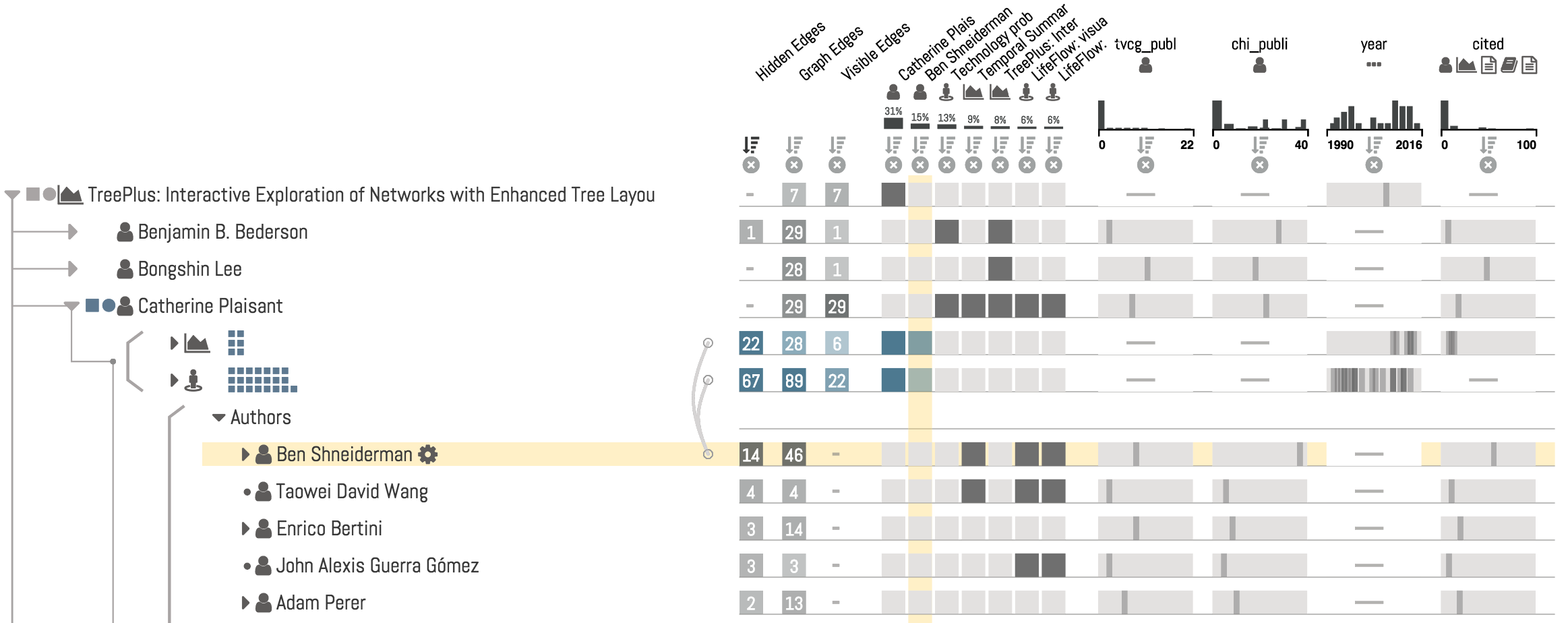} \label{fig:co-auth:plaisant_co-authors}}
  \vspace{-3mm}
  \caption{A use case for exploring the relationships between scholars and papers. (a) Distribution of papers of authors of the \textit{TreePlus} paper across \CHI~CHI and \TVCG~TVCG. These authors have published about five times as much at CHI than in TVCG. In particular, \textit{Ben Bederson} has no TVCG paper other than the TreePlus paper (which is the root), as is evident form the matrix and the missing link from his node to the TVCG aggregate. (b) Distribution of papers for individual authors. \textit{Bongshin Lee} has published most evenly between TVCG and CHI. (c) Frequent co-authors of \textit{Catherine Plaisant}. Authors are sorted by the number of hidden edges. \textit{Ben Shneiderman} is a frequent collaborator, but he has also published many papers with others. \textit{Taowei David Wang} has published all his papers with Catherine Plaisant.}
   \label{fig:co-auth}
  \vspace{-5mm}
\end{figure*}

We curated the co-author dataset introduced previously by retrieving a list of papers from DPLP\footnote{\url{https://dblp.uni-trier.de/}}. We extracted all papers published at ACM CHI and IEEE TVCG up to 2015. We have also included additional attributes about papers based on the visualization publication dataset compiled by Isenberg et al.~\cite{isenberg_vispubdata.org:_2017}. We used this information to also compile aggregate citation counts for authors. 

We start by querying for a paper that is relevant for this manuscript: the \textit{TreePlus} paper by Lee et al.~\cite{lee_treeplus:_2006-1}. By expanding its neighbors (Figure~\ref{fig:co-auth:grouped_papers}), we reveal that the paper has seven co-authors. Several of the authors are familiar names, but we would like to see which ones are the most prolific scholars. To answer this, we scan the \textit{graph edges} column, which corresponds to the number of papers these authors have published at CHI and TVCG combined. We see that \textit{Catherine Plaisant} and \textit{Ben Bederson} have published 29 papers each, and \textit{Bongshin Lee} has published 28. But does this group combined publish more at CHI or at TVCG? To answer that question, we expand all neighbors of these authors (their papers) and put the tree into level layout in order to group together all papers and to group them by type. However, as this combined list is quite long, we aggregate the papers, as shown in Figure~\ref{fig:co-auth:grouped_papers}. We see that, overall, there are about five times as many papers at \CHI~CHI than in \TVCG~TVCG for these authors. By looking at the adjacency matrix cell for the TVCG papers and Ben Bederson, we can see that he does not have a TVCG paper, other than TreePlus, but he is a very prolific author at CHI. We confirm this by hovering over his node, which reveals an edge to the CHI aggregate but not to TVCG.

Next, we focus on the other authors' papers. Have they published exclusively at CHI, in TVCG, or both? By switching from aggregated level to tree mode, we see that Catherine Plaisant and Bongshin Lee have published frequently at both CHI and TVCG, as shown in Figure~\ref{fig:co-auth:author_conf_dist}. We hone in on Catherine Plaisant's papers by gathering all her co-authored papers, and then we look at the aggregate information for the years she published (see Figure~\ref{fig:juniper_overview}). We see that she was immensely successful at CHI in the 90s and continued publishing at CHI afterwards, and that her TVCG publications start in 2008, soon after the first VIS publications appeared in TVCG. 

Now we want to learn about Catherine Plaisant's co-authors. Has she published with many different people, or does she have consistent collaboration partners? We set the branch starting with her to level mode, expand all of her children, and aggregate the papers. Next we sort her co-authors by the number of hidden edges --- those edges correspond to papers these authors have co-authored with Catherine Plaisant (Figure~\ref{fig:co-auth:plaisant_co-authors}). We quickly identify that she has collaborated extensively with Ben Shneiderman. She has written 14 papers at TVCG and CHI together with him. We also see that Ben Shneiderman has published many more papers at CHI (38) than in TVCG (8). We discover that her second-most frequent co-author is Taowei David Wang, who has published all of his four papers together with her. 

Cahterine Plaisant has also published with other prolific scholars, which we can identify by sorting the authors by citation. To clean the list up, we aggregate and set a DOI to show only authors with more than 15 citations (Figure~\ref{fig:juniper_overview}; note that these citation counts reflect only citation included in this dataset~\cite{isenberg_vispubdata.org:_2017}). We see that, in addition to Ben Shneiderman, \textit{Jean-Daniel Fekete}, \textit{Petra Isenberg}, \textit{Nathalie Henry Riche} and \textit{Heidi Lam} are in this list. But looking at the hidden edge column, we see that these have been co-authors on only one or two papers in our dataset.

\section{Discussion}
\label{sec:discussion}

Juniper is designed for the tree-based graph exploration of highly multivariate graphs. As discussed in Section~\ref{sec:tasks}, Juniper addresses focus tasks, such as those related to adjacency and paths, yet always in the context of the relationship between the topology and attributes of a network. 

As far as the analysis of topology is concerned, tree-based graph exploration has been shown by Lee et al.~\cite{lee_treeplus:_2006-1} to perform better than force-directed layouts for various tasks, including path-based and connectivity- based tasks. The same study also showed that tree-based graph exploration leads to significantly higher confidence and that users preferred the tree-based layout. 

Our goal in the development of Juniper was to (1) improve on the current state of the art in tree-based graph exploration, by providing novel visual encodings and interactions, such as topology/path-based sorting, DOI-based aggregation, attribute-driven sorting, and the combination of an adjacency matrix with the tree-layout, and (2) leverage the tree-based layout to visualize attributes, tightly integrated with topology. We argue that Juniper is well suited to address our key tasks: understanding attributes in the context of paths in a network and understanding attributes in the context of neighborhoods.

One of the downsides of using a tree-based layout is that it is difficult to understand cycles in a network. Although we believe that the visualization of hidden edges makes it possible to do that in Juniper, it is not the ideal solution because it requires interaction to uncover a cycle. We are considering various strategies to address this problem, including a supplemental view for cycles, a special encoding along the tree, or breaking with the tree-convention for selected nodes. 

Our technique targets focus tasks, yet overviews can be useful in some scenarios. Although we chose not to include an overview visualization to be concise and focused, we envision that a production-quality graph analysis system would also provide overview techniques as, for example, described by van den Elzen and van Wijk~\cite{vandenelzen_multivariate_2014}. In such an integrated tool, analysts could seamlessly switch between representations optimized for overview and focus tasks.

\subsection{Scalability}

Since Juniper is a bottom-up graph visualization technique, the size of the underlying graph is limited only by the capabilities and performance of the graph database. The largest network we currently include in our demo (the co-author network) has about 34,000 nodes and 90,000 edges and results in no noticeable delays for common queries. 

The scalability of the subgraph is limited by the number of rows that can be simultaneously displayed. On a large desktop screen, we can show about 50-60 rows. The number of rows, however, corresponds to the number of nodes only when no aggregation is used. We found that the use of aggregation combined with a DOI function is a very efficient way to explore subgraphs with a few hundred nodes, depending on the properties of the network. In cases where more rows are displayed than can be fit on the screen, we use scrolling. However, we currently do not provide a good solution for linking to off-screen content; hence working with many more rows can be tedious.

In terms of the number of attributes, Juniper is exceptionally scalable compared to other multivariate network visualization techniques. We consciously reserve a sizable portion of the available screen for making long node labels, such as paper titles, readable. Even with that much space dedicated to labels, we can display 10-20 additional attribute columns on a desktop display. Our current visual encodings for attribute visualization favor precision and details over compactness; more compact representations, such as those used in enRoute~\cite{partl_enroute:_2012} are conceivable and could increase that number considerably. 

\subsection{Comparison to Related Techniques}

We compare Juniper to TreePlus~\cite{lee_treeplus:_2006-1}, since TreePlus is the most comprehensive of all the tree-based graph visualization techniques discussed in Section~\ref{sec:rw_tree}. Although TreePlus shares the basic idea of tree-based graph exploration with Juniper and was an important inspiration for our work, Juniper introduces several novel concepts that go significantly beyond the capabilities of TreePlus. The \textbf{key distinction from TreePlus is our ability to visualize rich attribute data}, due to the linearized layout and the juxtaposition with the table. However, we also argue that Juniper is at least competitive with TreePlus when considering only topological tasks. 
Even though the linear layout needs more space than the layout chosen in TreePlus, we argue that our aggregation methods counteract the increased spatial demand of the linear layout, and that our DOI-based deaggregation is effective at revealing relevant nodes and connections even when aggregation is used. TreePlus also does not have a level layout, which can be used to quickly identify nodes at a certain distance from the root.




Lineage~\cite{nobre_lineage:_2018} is a domain-specific tool developed for visualizing clinical genealogies and was published recently by some authors of this manuscript. Although \textbf{Lineage} shares the idea of using a linearized tree to visualize multivariate attributes of that tree in a table, it \textbf{is in fact a tree visualization tool, and not a general graph visualization technique}, like Juniper. The genealogies that can be visualized in Lineage have to be tree-like, i.e., have to trace back to a single founder. Rare cross-links within the tree are removed by duplicating the nodes in a preprocessing step. Since Lineage visualizes trees, it has no notion of re-shaping a tree based on a graph, does not show topological context by combined a node link and a matrix layout, and can represent only static instances of a tree, instead of growing a tree from a subgraph that is dynamically extracted from a large graph. Lineage does not support path-linearization and has no level mode. Lineage is an important tool in its niche application area. Juniper, in contrast, is a general purpose multivariate graph visualization technique with the potential for applications in many domains.

\subsection{Evaluation}

We considered various strategies to evaluate Juniper against the claims we make: that it is a well-suited technique for focus tasks in multivariate network analysis. We considered qualitative/usability evaluation, case studies, and insight-based evaluation, which we have rejected for different reasons. Ultimately, we decided between (1) quantitative evaluation of task performance (time, correctness) and (2) evaluation by justifying the design rationale and providing usage scenarios.  

With regard to quantitative evaluation, the key choices to make in the study design are (1) the tasks to use in the evaluation and (2) the comparison target. 
A comparison to a state-of-the-art system like Cytoscape introduces confounders and makes comparisons between a tool specifically designed for certain tasks with a general purpose tool. The best approach to quantitatively evaluate the core contribution of a complex technique such as Juniper would be to implement a reasonable alternative in the same general framework. We could compare Juniper to an MCV system using our attribute table and a force-directed layout. However, while we provide a simple force-directed layout for illustration in our prototype implementation, we do not provide advanced features such as aggregation, expanding or collapsing branches, etc. Since we have no canonical way to implement these features, this approach would require significant effort in designing such a system. It would also introduce additional potential confounders, because it matters how (well) these features are implemented.

We instead opt for evaluation by providing a detailed rationale for our design~\cite{greenberg_usability_2008} and demonstrate its usefulness in use cases. In line with recent successful papers (e.g.~\cite{holten_hierarchical_2006, vandenelzen_multivariate_2014, vandenelzen_reducing_2016, lex_upset:_2014}), we believe that design arguments and demonstrations through use cases provide excellent evidence for the utility of complex, interactive visualization techniques. 

\section{Conclusion and Future Work}

We believe that Juniper is widely applicable to different graph datasets across various domains. Juniper's strengths are in interactive exploration and in supporting focus tasks for multivariate graph visualization. In the future we hope to develop techniques for connecting to off-screen nodes, which is a current scalability limitation. Also, allowing duplicate nodes could be helpful for certain tasks, yet node duplication requires careful encoding to not confuse users. Juniper currently also does not support rich edge attributes. Edge attributes of visible edges could be seamlessly integrated into Juniper, as the tree structure guarantees that each node has exactly one incoming edge. As a result, we could add rows representing edges to the table right above the destination node and add columns for the edge attributes. A drawback to this solution is that it cannot be used for hidden edges. Attributes of multiple hidden edges could be shown as extra columns in the table with their destination node, visualizing attributes from multiple edge in each cell. 
Finally, for graphs with many different node types, the attribute table can be sparse. We plan to investigate interleaving cells for different attributes in a single column to remedy this.

\acknowledgments{
The authors wish to thank members of the Visualization Design Lab for their feedback, and acknowledge support by NIH (U01 CA198935), NSF (IIS 1751238), DoD (ST1605-16-01), and the Austrian Science Fund (FWF P27975-NBL).}

\bibliographystyle{abbrv-doi}

\bibliography{2018_infovis_juniper}

\begin{thebibliography}{10}

\bibitem{barsky_cerebral:_2008}
A.~Barsky, T.~Munzner, J.~Gardy, and R.~Kincaid.
\newblock Cerebral: {{Visualizing Multiple Experimental Conditions}} on a
  {{Graph}} with {{Biological Context}}.
\newblock {\em IEEE Transactions on Visualization and Computer Graphics
  (InfoVis '08)}, 14(6):1253--1260, 2008. doi: {{%
10\hspace{.1pt}\discretionary{.}{%
}{.}\hspace{.4pt}1109\discretionary{/}{%
}{/}TVCG\hspace{.1pt}\discretionary{.}{%
}{.}\hspace{.4pt}2008\hspace{.1pt}\discretionary{.}{%
}{.}\hspace{.4pt}117}}


\bibitem{bastian_gephi:_2009}
M.~Bastian, S.~Heymann, and M.~Jacomy.
\newblock Gephi: {{An Open Source Software}} for {{Exploring}} and
  {{Manipulating Networks}}.
\newblock In {\em Third {{International AAAI Conference}} on {{Weblogs}} and
  {{Social Media}}}, Mar. 2009.

\bibitem{bezerianos_graphdice:_2010}
A.~Bezerianos, F.~Chevalier, P.~Dragicevic, N.~Elmqvist, and J.~D. Fekete.
\newblock {{GraphDice}}: {{A System}} for {{Exploring Multivariate Social
  Networks}}.
\newblock {\em Computer Graphics Forum (EuroVis '10)}, 29(3):863--872, 2010.
  doi: {{%
10\hspace{.1pt}\discretionary{.}{%
}{.}\hspace{.4pt}1111\discretionary{/}{%
}{/}j\hspace{.1pt}\discretionary{.}{%
}{.}\hspace{.4pt}1467\discretionary{%
}{-}{-}8659\hspace{.1pt}\discretionary{.}{%
}{.}\hspace{.4pt}2009\hspace{.1pt}\discretionary{.}{%
}{.}\hspace{.4pt}01687\hspace{.1pt}\discretionary{.}{%
}{.}\hspace{.4pt}x}}


\bibitem{burch_timeline_2008}
M.~Burch, F.~Beck, and S.~Diehl.
\newblock Timeline {{Trees}}: {{Visualizing Sequences}} of {{Transactions}} in
  {{Information Hierarchies}}.
\newblock In {\em Proceedings of the {{Working Conference}} on {{Advanced
  Visual Interfaces}}}, AVI '08, pp. 75--82. {ACM}, New York, NY, USA, 2008.
  doi: {{%
10\hspace{.1pt}\discretionary{.}{%
}{.}\hspace{.4pt}1145\discretionary{/}{%
}{/}1385569\hspace{.1pt}\discretionary{.}{%
}{.}\hspace{.4pt}1385584}}


\bibitem{christakis_spread_2007}
N.~A. Christakis and J.~H. Fowler.
\newblock The {{Spread}} of {{Obesity}} in a {{Large Social Network}} over 32
  {{Years}}.
\newblock {\em New England Journal of Medicine}, 357(4):370--379, July 2007.
  doi: {{%
10\hspace{.1pt}\discretionary{.}{%
}{.}\hspace{.4pt}1056\discretionary{/}{%
}{/}NEJMsa066082}}


\bibitem{dunne_graphtrail:_2012}
C.~Dunne, N.~Henry~Riche, B.~Lee, R.~Metoyer, and G.~Robertson.
\newblock {{GraphTrail}}: {{Analyzing Large Multivariate}}, {{Heterogeneous
  Networks While Supporting Exploration History}}.
\newblock In {\em Proceedings of the {{ACM SIGCHI Conference}} on {{Human
  Factors}} in {{Computing Systems}} ({{CHI}} '12)}, pp. 1663--1672. {ACM},
  2012. doi: {{%
10\hspace{.1pt}\discretionary{.}{%
}{.}\hspace{.4pt}1145\discretionary{/}{%
}{/}2207676\hspace{.1pt}\discretionary{.}{%
}{.}\hspace{.4pt}2208293}}


\bibitem{eisen_cluster_1998}
M.~B. Eisen, P.~T. Spellman, P.~O. Brown, and D.~Botstein.
\newblock Cluster analysis and display of genome-wide expression patterns.
\newblock {\em Proceedings of the National Academy of Sciences USA},
  95(25):14863--14868, 1998. doi: {{%
10\hspace{.1pt}\discretionary{.}{%
}{.}\hspace{.4pt}1073\discretionary{/}{%
}{/}pnas\hspace{.1pt}\discretionary{.}{%
}{.}\hspace{.4pt}95\hspace{.1pt}\discretionary{.}{%
}{.}\hspace{.4pt}25\hspace{.1pt}\discretionary{.}{%
}{.}\hspace{.4pt}14863}}


\bibitem{eklund_ontorama:_2002}
P.~Eklund, N.~Roberts, and S.~Green.
\newblock {{OntoRama}}: {{Browsing RDF}} ontologies using a hyperbolic-style
  browser.
\newblock In {\em First {{International Symposium}} on {{Cyber Worlds}}, 2002.
  {{Proceedings}}.}, pp. 405--411, 2002. doi: {{%
10\hspace{.1pt}\discretionary{.}{%
}{.}\hspace{.4pt}1109\discretionary{/}{%
}{/}CW\hspace{.1pt}\discretionary{.}{%
}{.}\hspace{.4pt}2002\hspace{.1pt}\discretionary{.}{%
}{.}\hspace{.4pt}1180907}}


\bibitem{elmqvist_zame:_2008}
N.~Elmqvist, T.-N. Do, H.~Goodell, N.~Henry, and J.~Fekete.
\newblock {{ZAME}}: {{Interactive Large}}-{{Scale Graph Visualization}}.
\newblock In {\em Visualization {{Symposium}}, 2008. {{PacificVIS}} '08. {{IEEE
  Pacific}}}, pp. 215--222, 2008. doi: {{%
10\hspace{.1pt}\discretionary{.}{%
}{.}\hspace{.4pt}1109\discretionary{/}{%
}{/}PACIFICVIS\hspace{.1pt}\discretionary{.}{%
}{.}\hspace{.4pt}2008\hspace{.1pt}\discretionary{.}{%
}{.}\hspace{.4pt}4475479}}


\bibitem{fekete_interactive_2003}
J.-D. Fekete, D.~Wang, N.~Dang, A.~Aris, and C.~Plaisant.
\newblock Interactive {{Poster}}: {{Overlaying Graph Links}} on {{Treemaps}}.
\newblock In {\em Proceedings of the {{IEEE Symposium}} on {{Information
  Visualization}} ({{InfoVis}} '03)}, pp. 82--83. {IEEE}, 2003.

\bibitem{furmanova_taggle:_2018}
K.~Furmanova, S.~Gratzl, H.~Stitz, T.~Zichner, M.~Jaresova, M.~Ennemoser,
  A.~Lex, and M.~Streit.
\newblock Taggle: {{Scalable Visualization}} of {{Tabular Data}} through
  {{Aggregation}}.
\newblock {\em arXiv preprint}, 2018.

\bibitem{furnas_generalized_1986}
G.~W. Furnas.
\newblock Generalized fisheye views.
\newblock In {\em Proceedings of the {{SIGCHI Conference}} on {{Human Factors}}
  in {{Computing Systems}} ({{CHI}} '86)}, pp. 16--23. {ACM}, 1986. doi: {{%
10\hspace{.1pt}\discretionary{.}{%
}{.}\hspace{.4pt}1145\discretionary{/}{%
}{/}22339\hspace{.1pt}\discretionary{.}{%
}{.}\hspace{.4pt}22342}}


\bibitem{gehlenborg_visualization_2010}
N.~Gehlenborg, S.~I. O'Donoghue, N.~S. Baliga, A.~Goesmann, M.~A. Hibbs,
  H.~Kitano, O.~Kohlbacher, H.~Neuweger, R.~Schneider, D.~Tenenbaum, and A.-C.
  Gavin.
\newblock Visualization of omics data for systems biology.
\newblock {\em Nature Methods}, 7(3):56--68, 2010. doi: {{%
10\hspace{.1pt}\discretionary{.}{%
}{.}\hspace{.4pt}1038\discretionary{/}{%
}{/}nmeth\hspace{.1pt}\discretionary{.}{%
}{.}\hspace{.4pt}1436}}


\bibitem{ghoniem_readability_2005}
M.~Ghoniem, J.-D. Fekete, and P.~Castagliola.
\newblock On the {{Readability}} of {{Graphs Using Node}}-{{Link}} and
  {{Matrix}}-{{Based Representations}}: {{A Controlled Experiment}} and
  {{Statistical Analysis}}.
\newblock {\em Information Visualization}, 4(2):114 --135, June 2005. doi: {{%
10\hspace{.1pt}\discretionary{.}{%
}{.}\hspace{.4pt}1057\discretionary{/}{%
}{/}palgrave\hspace{.1pt}\discretionary{.}{%
}{.}\hspace{.4pt}ivs\hspace{.1pt}\discretionary{.}{%
}{.}\hspace{.4pt}9500092}}


\bibitem{gou_treenetviz:_2011}
L.~Gou and X.~Zhang.
\newblock Treenetviz: {{Revealing}} patterns of networks over tree structures.
\newblock {\em IEEE Transactions on Visualization and Computer Graphics
  (InfoVis '11)}, 17(12):2449--2458, 2011.

\bibitem{greenberg_usability_2008}
S.~Greenberg and B.~Buxton.
\newblock Usability {{Evaluation Considered Harmful}} ({{Some}} of the
  {{Time}}).
\newblock In {\em Proceedings of the {{SIGCHI Conference}} on {{Human Factors}}
  in {{Computing Systems}}}, CHI '08, pp. 111--120. {ACM}, New York, NY, USA,
  2008. doi: {{%
10\hspace{.1pt}\discretionary{.}{%
}{.}\hspace{.4pt}1145\discretionary{/}{%
}{/}1357054\hspace{.1pt}\discretionary{.}{%
}{.}\hspace{.4pt}1357074}}


\bibitem{hao_web-based_2000}
M.~C. Hao, M.~Hsu, U.~Dayal, and A.~Krug.
\newblock Web-{{Based Visualization}} of {{Large Hierarchical Graphs Using
  Invisible Links}} in a {{Hyperbolic Space}}.
\newblock In {\em Advances in {{Visual Information Management}}}, IFIP
  \textemdash{} The International Federation for Information Processing, pp.
  83--94. {Springer, Boston, MA}, 2000. doi: {{%
10\hspace{.1pt}\discretionary{.}{%
}{.}\hspace{.4pt}1007\discretionary{/}{%
}{/}978\discretionary{%
}{-}{-}0\discretionary{%
}{-}{-}387\discretionary{%
}{-}{-}35504\discretionary{%
}{-}{-}7\_6}}


\bibitem{heer_vizster:_2005}
J.~Heer and D.~Boyd.
\newblock Vizster: Visualizing online social networks.
\newblock In {\em {{IEEE Symposium}} on {{Information Visualization}}, 2005.
  {{INFOVIS}} 2005}, pp. 32--39, Oct. 2005. doi: {{%
10\hspace{.1pt}\discretionary{.}{%
}{.}\hspace{.4pt}1109\discretionary{/}{%
}{/}INFVIS\hspace{.1pt}\discretionary{.}{%
}{.}\hspace{.4pt}2005\hspace{.1pt}\discretionary{.}{%
}{.}\hspace{.4pt}1532126}}


\bibitem{henry_matlink:_2007}
N.~Henry and J.-D. Fekete.
\newblock {{MatLink}}: {{Enhanced Matrix Visualization}} for {{Analyzing Social
  Networks}}.
\newblock In C.~Baranauskas, P.~Palanque, J.~Abascal, and S.~D.~J. Barbosa,
  eds., {\em Human-{{Computer Interaction}} \textendash{} {{INTERACT}} 2007},
  number 4663 in Lecture Notes in Computer Science, pp. 288--302. {Springer
  Berlin Heidelberg}, 2007.

\bibitem{henry_nodetrix:_2007}
N.~Henry, J.~D. Fekete, and M.~J. McGuffin.
\newblock {{NodeTrix}}: A {{Hybrid Visualization}} of {{Social Networks}}.
\newblock {\em IEEE Transactions on Visualization and Computer Graphics
  (InfoVis '07)}, 13(6):1302--1309, 2007. doi: {{%
10\hspace{.1pt}\discretionary{.}{%
}{.}\hspace{.4pt}1109\discretionary{/}{%
}{/}TVCG\hspace{.1pt}\discretionary{.}{%
}{.}\hspace{.4pt}2007\hspace{.1pt}\discretionary{.}{%
}{.}\hspace{.4pt}70582}}


\bibitem{holten_hierarchical_2006}
D.~Holten.
\newblock Hierarchical {{Edge Bundles}}: {{Visualization}} of {{Adjacency
  Relations}} in {{Hierarchical Data}}.
\newblock {\em IEEE Transactions on Visualization and Computer Graphics
  (InfoVis '06)}, 12(5):741--748, 2006. doi: {{%
10\hspace{.1pt}\discretionary{.}{%
}{.}\hspace{.4pt}1109\discretionary{/}{%
}{/}TVCG\hspace{.1pt}\discretionary{.}{%
}{.}\hspace{.4pt}2006\hspace{.1pt}\discretionary{.}{%
}{.}\hspace{.4pt}147}}


\bibitem{isenberg_vispubdata.org:_2017}
P.~Isenberg, F.~Heimerl, S.~Koch, T.~Isenberg, P.~Xu, C.~D. Stolper,
  M.~Sedlmair, J.~Chen, T.~M{\"o}ller, and J.~Stasko.
\newblock Vispubdata.org: {{A Metadata Collection About IEEE Visualization}}
  ({{VIS}}) {{Publications}}.
\newblock {\em IEEE Transactions on Visualization and Computer Graphics},
  23(9):2199--2206, Sept. 2017. doi: {{%
10\hspace{.1pt}\discretionary{.}{%
}{.}\hspace{.4pt}1109\discretionary{/}{%
}{/}TVCG\hspace{.1pt}\discretionary{.}{%
}{.}\hspace{.4pt}2016\hspace{.1pt}\discretionary{.}{%
}{.}\hspace{.4pt}2615308}}


\bibitem{jankun-kelly_moiregraphs:_2003}
T.~J. {Jankun-Kelly} and K.-L. Ma.
\newblock {{MoireGraphs}}: Radial focus+context visualization and interaction
  for graphs with visual nodes.
\newblock In {\em {{IEEE Symposium}} on {{Information Visualization}} 2003
  ({{IEEE Cat}}. {{No}}.{{03TH8714}})}, pp. 59--66, Oct. 2003. doi: {{%
10\hspace{.1pt}\discretionary{.}{%
}{.}\hspace{.4pt}1109\discretionary{/}{%
}{/}INFVIS\hspace{.1pt}\discretionary{.}{%
}{.}\hspace{.4pt}2003\hspace{.1pt}\discretionary{.}{%
}{.}\hspace{.4pt}1249009}}


\bibitem{johnson_tree-maps:_1991}
B.~Johnson and B.~Shneiderman.
\newblock Tree-maps: A space-filling approach to the visualization of
  hierarchical information structures.
\newblock In {\em Proceedings of the {{IEEE Conference}} on {{Visualization}}
  ({{Vis}} '91)}, pp. 284--291, 1991. doi: {{%
10\hspace{.1pt}\discretionary{.}{%
}{.}\hspace{.4pt}1109\discretionary{/}{%
}{/}VISUAL\hspace{.1pt}\discretionary{.}{%
}{.}\hspace{.4pt}1991\hspace{.1pt}\discretionary{.}{%
}{.}\hspace{.4pt}175815}}


\bibitem{kairam_refinery:_2015-1}
S.~Kairam, N.~H. Riche, S.~Drucker, R.~Fernandez, and J.~Heer.
\newblock Refinery: {{Visual Exploration}} of {{Large}}, {{Heterogeneous
  Networks Through Associative Browsing}}.
\newblock {\em Computer Graphics Forum (EuroVis '15)}, 34:301--310, 2015.

\bibitem{kerren_multivariate_2014}
A.~Kerren, H.~C. Purchase, and M.~Ward, eds.
\newblock {\em Multivariate {{Network Visualization}}}.
\newblock Number 8380 in Lecture notes in computer science. {Springer}, 2014.

\bibitem{kerzner_graffinity:_2017}
E.~Kerzner, A.~Lex, C.~L. Sigulinsky, R.~E. Marc, B.~W. Jones, T.~Urness, and
  M.~Meyer.
\newblock Graffinity: {{Visualizing Connectivity}} in {{Large Graphs}}.
\newblock {\em Computer Graphics Forum (EuroVis '17)}, 36(3):251--260, 2017.

\bibitem{kreft_phyd3:_2017}
L.~Kreft, A.~Botzki, F.~Coppens, K.~Vandepoele, and M.~Van~Bel.
\newblock {{PhyD3}}: A phylogenetic tree viewer with extended {{phyloXML}}
  support for functional genomics data visualization.
\newblock {\em Bioinformatics}, 33(18):2946--2947, Sept. 2017. doi: {{%
10\hspace{.1pt}\discretionary{.}{%
}{.}\hspace{.4pt}1093\discretionary{/}{%
}{/}bioinformatics\discretionary{/}{%
}{/}btx324}}


\bibitem{kruskal_icicle_1983}
J.~B. Kruskal and J.~M. Landwehr.
\newblock Icicle {{Plots}}: {{Better Displays}} for {{Hierarchical
  Clustering}}.
\newblock {\em The American Statistician}, 37(2):162, 1983. doi: {{%
10\hspace{.1pt}\discretionary{.}{%
}{.}\hspace{.4pt}2307\discretionary{/}{%
}{/}2685881}}


\bibitem{lee_gotreeplus:_2008}
B.~Lee, K.~Brown, Y.~Hathout, and J.~Seo.
\newblock {{GOTreePlus}}: An interactive gene ontology browser.
\newblock {\em Bioinformatics}, 24(7):1026--1028, Apr. 2008. doi: {{%
10\hspace{.1pt}\discretionary{.}{%
}{.}\hspace{.4pt}1093\discretionary{/}{%
}{/}bioinformatics\discretionary{/}{%
}{/}btn068}}


\bibitem{lee_phylodet:_2009}
B.~Lee, L.~Nachmanson, G.~Robertson, J.~M. Carlson, and D.~Heckerman.
\newblock {{PhyloDet}}: A scalable visualization tool for mapping multiple
  traits to large evolutionary trees.
\newblock {\em Bioinformatics}, 25(19):2611--2612, 2009. doi: {{%
10\hspace{.1pt}\discretionary{.}{%
}{.}\hspace{.4pt}1093\discretionary{/}{%
}{/}bioinformatics\discretionary{/}{%
}{/}btp454}}


\bibitem{lee_treeplus:_2006-1}
B.~Lee, C.~S. Parr, C.~Plaisant, B.~B. Bederson, V.~D. Veksler, W.~D. Gray, and
  C.~Kotfila.
\newblock {{TreePlus}}: {{Interactive Exploration}} of {{Networks}} with
  {{Enhanced Tree Layouts}}.
\newblock {\em IEEE Transactions on Visualization and Computer Graphics},
  12(6):1414--1426, Nov. 2006. doi: {{%
10\hspace{.1pt}\discretionary{.}{%
}{.}\hspace{.4pt}1109\discretionary{/}{%
}{/}TVCG\hspace{.1pt}\discretionary{.}{%
}{.}\hspace{.4pt}2006\hspace{.1pt}\discretionary{.}{%
}{.}\hspace{.4pt}106}}


\bibitem{lee_task_2006}
B.~Lee, C.~Plaisant, C.~S. Parr, J.-D. Fekete, and N.~Henry.
\newblock Task {{Taxonomy}} for {{Graph Visualization}}.
\newblock In {\em Proceedings of the {{AVI Workshop}} on {{BEyond}} Time and
  Errors: Novel Evaluation Methods for Information Visualization ({{BELIV}}
  '06)}, pp. 1--5, 2006. doi: {{%
10\hspace{.1pt}\discretionary{.}{%
}{.}\hspace{.4pt}1145\discretionary{/}{%
}{/}1168149\hspace{.1pt}\discretionary{.}{%
}{.}\hspace{.4pt}1168168}}


\bibitem{lex_upset:_2014}
A.~Lex, N.~Gehlenborg, H.~Strobelt, R.~Vuillemot, and H.~Pfister.
\newblock {{UpSet}}: {{Visualization}} of {{Intersecting Sets}}.
\newblock {\em IEEE Transactions on Visualization and Computer Graphics
  (InfoVis '14)}, 20(12):1983--1992, 2014. doi: {{%
10\hspace{.1pt}\discretionary{.}{%
}{.}\hspace{.4pt}1109\discretionary{/}{%
}{/}TVCG\hspace{.1pt}\discretionary{.}{%
}{.}\hspace{.4pt}2014\hspace{.1pt}\discretionary{.}{%
}{.}\hspace{.4pt}2346248}}


\bibitem{lex_caleydo:_2010}
A.~Lex, M.~Streit, E.~Kruijff, and D.~Schmalstieg.
\newblock Caleydo: {{Design}} and {{Evaluation}} of a {{Visual Analysis
  Framework}} for {{Gene Expression Data}} in its {{Biological Context}}.
\newblock In {\em Proceedings of the {{IEEE Symposium}} on {{Pacific
  Visualization}} ({{PacificVis}} '10)}, pp. 57--64. {IEEE}, 2010. doi: {{%
10\hspace{.1pt}\discretionary{.}{%
}{.}\hspace{.4pt}1109\discretionary{/}{%
}{/}PACIFICVIS\hspace{.1pt}\discretionary{.}{%
}{.}\hspace{.4pt}2010\hspace{.1pt}\discretionary{.}{%
}{.}\hspace{.4pt}5429609}}


\bibitem{lex_stratomex:_2012-1}
A.~Lex, M.~Streit, H.-J. Schulz, C.~Partl, D.~Schmalstieg, P.~J. Park, and
  N.~Gehlenborg.
\newblock {{StratomeX}}: {{Visual Analysis}} of {{Large}}-{{Scale Heterogeneous
  Genomics Data}} for {{Cancer Subtype Characterization}}.
\newblock {\em Computer Graphics Forum (EuroVis '12)}, 31(3):1175--1184, 2012.
  doi: {{%
10\hspace{.1pt}\discretionary{.}{%
}{.}\hspace{.4pt}1111\discretionary{/}{%
}{/}j\hspace{.1pt}\discretionary{.}{%
}{.}\hspace{.4pt}1467\discretionary{%
}{-}{-}8659\hspace{.1pt}\discretionary{.}{%
}{.}\hspace{.4pt}2012\hspace{.1pt}\discretionary{.}{%
}{.}\hspace{.4pt}03110\hspace{.1pt}\discretionary{.}{%
}{.}\hspace{.4pt}x}}


\bibitem{meyer_pathline:_2010}
M.~Meyer, B.~Wong, M.~Styczynski, T.~Munzner, and H.~Pfister.
\newblock Pathline: {{A Tool For Comparative Functional Genomics}}.
\newblock {\em Computer Graphics Forum (EuroVis '10)}, 29(3):1043--1052, 2010.
  doi: {{%
10\hspace{.1pt}\discretionary{.}{%
}{.}\hspace{.4pt}1111\discretionary{/}{%
}{/}j\hspace{.1pt}\discretionary{.}{%
}{.}\hspace{.4pt}1467\discretionary{%
}{-}{-}8659\hspace{.1pt}\discretionary{.}{%
}{.}\hspace{.4pt}2009\hspace{.1pt}\discretionary{.}{%
}{.}\hspace{.4pt}01710\hspace{.1pt}\discretionary{.}{%
}{.}\hspace{.4pt}x}}


\bibitem{munzner_drawing_1998}
T.~Munzner.
\newblock Drawing {{Large Graphs}} with {{H3Viewer}} and {{Site Manager}}.
\newblock In {\em Graph {{Drawing}}}, Lecture Notes in Computer Science, pp.
  384--393. {Springer, Berlin, Heidelberg}, Aug. 1998. doi: {{%
10\hspace{.1pt}\discretionary{.}{%
}{.}\hspace{.4pt}1007\discretionary{/}{%
}{/}3\discretionary{%
}{-}{-}540\discretionary{%
}{-}{-}37623\discretionary{%
}{-}{-}2\_30}}


\bibitem{nobre_lineage:_2018}
C.~Nobre, N.~Gehlenborg, H.~Coon, and A.~Lex.
\newblock Lineage: {{Visualizing Multivariate Clinical Data}} in {{Genealogy
  Graphs}}.
\newblock {\em Transaction on Visualization and Computer Graphics}, PP, 2018.
\newblock to appear. doi: {{%
10\hspace{.1pt}\discretionary{.}{%
}{.}\hspace{.4pt}1109\discretionary{/}{%
}{/}TVCG\hspace{.1pt}\discretionary{.}{%
}{.}\hspace{.4pt}2018\hspace{.1pt}\discretionary{.}{%
}{.}\hspace{.4pt}2811488}}


\bibitem{partl_pathfinder:_2016}
C.~Partl, S.~Gratzl, M.~Streit, A.~M. Wassermann, H.~Pfister, D.~Schmalstieg,
  and A.~Lex.
\newblock Pathfinder: {{Visual Analysis}} of {{Paths}} in {{Graphs}}.
\newblock {\em Computer Graphics Forum (EuroVis '16)}, 35(3):71--80, 2016. doi:
  {{%
10\hspace{.1pt}\discretionary{.}{%
}{.}\hspace{.4pt}1111\discretionary{/}{%
}{/}cgf\hspace{.1pt}\discretionary{.}{%
}{.}\hspace{.4pt}12883}}


\bibitem{partl_enroute:_2012}
C.~Partl, A.~Lex, M.~Streit, D.~Kalkofen, K.~Kashofer, and D.~Schmalstieg.
\newblock {{enRoute}}: {{Dynamic Path Extraction}} from {{Biological Pathway
  Maps}} for {{In}}-{{Depth Experimental Data Analysis}}.
\newblock In {\em Proceedings of the {{IEEE Symposium}} on {{Biological Data
  Visualization}} ({{BioVis}} '12)}, pp. 107--114, 2012. doi: {{%
10\hspace{.1pt}\discretionary{.}{%
}{.}\hspace{.4pt}1109\discretionary{/}{%
}{/}BioVis\hspace{.1pt}\discretionary{.}{%
}{.}\hspace{.4pt}2012\hspace{.1pt}\discretionary{.}{%
}{.}\hspace{.4pt}6378600}}


\bibitem{pienta_vigor:_2018}
R.~Pienta, F.~Hohman, A.~Endert, A.~Tamersoy, K.~Roundy, C.~Gates, S.~Navathe,
  and D.~H. Chau.
\newblock {{VIGOR}}: {{Interactive Visual Exploration}} of {{Graph Query
  Results}}.
\newblock {\em IEEE Transactions on Visualization and Computer Graphics},
  24(1):215--225, Jan. 2018. doi: {{%
10\hspace{.1pt}\discretionary{.}{%
}{.}\hspace{.4pt}1109\discretionary{/}{%
}{/}TVCG\hspace{.1pt}\discretionary{.}{%
}{.}\hspace{.4pt}2017\hspace{.1pt}\discretionary{.}{%
}{.}\hspace{.4pt}2744898}}


\bibitem{shannon_multivariate_2008}
R.~Shannon, T.~Holland, and A.~Quigley.
\newblock Multivariate {{Graph Drawing}} using {{Parallel Coordinate
  Visualisations}}.
\newblock Technical report, {University of St Andrews}, 2008.

\bibitem{smoot_cytoscape_2011}
M.~E. Smoot, K.~Ono, J.~Ruscheinski, P.-L. Wang, and T.~Ideker.
\newblock Cytoscape 2.8: New features for data integration and network
  visualization.
\newblock {\em Bioinformatics}, 27(3):431--432, Jan. 2011. doi: {{%
10\hspace{.1pt}\discretionary{.}{%
}{.}\hspace{.4pt}1093\discretionary{/}{%
}{/}bioinformatics\discretionary{/}{%
}{/}btq675}}


\bibitem{stasko_focus+context_2000}
J.~Stasko and E.~Zhang.
\newblock Focus+{{Context Display}} and {{Navigation Techniques}} for
  {{Enhancing Radial}}, {{Space}}-{{Filling Hierarchy Visualizations}}.
\newblock In {\em Proceedings of the {{IEEE Symposium}} on {{Information
  Vizualization}} ({{InfoVis}} '00)}, pp. 57--65. {IEEE Computer Society
  Press}, 2000. doi: {{%
10\hspace{.1pt}\discretionary{.}{%
}{.}\hspace{.4pt}1109\discretionary{/}{%
}{/}INFVIS\hspace{.1pt}\discretionary{.}{%
}{.}\hspace{.4pt}2000\hspace{.1pt}\discretionary{.}{%
}{.}\hspace{.4pt}885091}}


\bibitem{tu_graphcharter:_2013}
Y.~Tu and H.~W. Shen.
\newblock {{GraphCharter}}: {{Combining}} browsing with query to explore large
  semantic graphs.
\newblock In {\em 2013 {{IEEE Pacific Visualization Symposium}}
  ({{PacificVis}})}, pp. 49--56, Feb. 2013. doi: {{%
10\hspace{.1pt}\discretionary{.}{%
}{.}\hspace{.4pt}1109\discretionary{/}{%
}{/}PacificVis\hspace{.1pt}\discretionary{.}{%
}{.}\hspace{.4pt}2013\hspace{.1pt}\discretionary{.}{%
}{.}\hspace{.4pt}6596127}}


\bibitem{vandenelzen_reducing_2016}
S.~{van den Elzen}, D.~Holten, J.~Blaas, and J.~{van Wijk}.
\newblock Reducing {{Snapshots}} to {{Points}}: {{A Visual Analytics Approach}}
  to {{Dynamic Network Exploration}}.
\newblock {\em IEEE Transactions on Visualization and Computer Graphics},
  22(1):1--10, Jan. 2016. doi: {{%
10\hspace{.1pt}\discretionary{.}{%
}{.}\hspace{.4pt}1109\discretionary{/}{%
}{/}TVCG\hspace{.1pt}\discretionary{.}{%
}{.}\hspace{.4pt}2015\hspace{.1pt}\discretionary{.}{%
}{.}\hspace{.4pt}2468078}}


\bibitem{vandenelzen_multivariate_2014}
S.~{van den Elzen} and J.~{van Wijk}.
\newblock Multivariate {{Network Exploration}} and {{Presentation}}: {{From
  Detail}} to {{Overview}} via {{Selections}} and {{Aggregations}}.
\newblock {\em IEEE Transactions on Visualization and Computer Graphics
  (InfoVis '14)}, 20(12):2310--2319, 2014. doi: {{%
10\hspace{.1pt}\discretionary{.}{%
}{.}\hspace{.4pt}1109\discretionary{/}{%
}{/}TVCG\hspace{.1pt}\discretionary{.}{%
}{.}\hspace{.4pt}2014\hspace{.1pt}\discretionary{.}{%
}{.}\hspace{.4pt}2346441}}


\bibitem{vanham_search_2009}
F.~{van Ham} and A.~Perer.
\newblock Search, {{Show Context}}, {{Expand}} on {{Demand}}: {{Supporting
  Large Graph Exploration}} with {{Degree}}-of-{{Interest}}.
\newblock {\em IEEE Transactions on Visualization and Computer Graphics
  (InfoVis '09)}, 15(6):953--960, 2009. doi: {{%
10\hspace{.1pt}\discretionary{.}{%
}{.}\hspace{.4pt}1109\discretionary{/}{%
}{/}TVCG\hspace{.1pt}\discretionary{.}{%
}{.}\hspace{.4pt}2009\hspace{.1pt}\discretionary{.}{%
}{.}\hspace{.4pt}108}}


\bibitem{vonlandesberger_visual_2011}
T.~{von Landesberger}, A.~Kuijper, T.~Schreck, J.~Kohlhammer, J.~{van Wijk},
  J.-D. Fekete, and D.~Fellner.
\newblock Visual {{Analysis}} of {{Large Graphs}}: {{State}}-of-the-{{Art}} and
  {{Future Research Challenges}}.
\newblock {\em Computer Graphics Forum}, 30(6):1719--1749, 2011. doi: {{%
10\hspace{.1pt}\discretionary{.}{%
}{.}\hspace{.4pt}1111\discretionary{/}{%
}{/}j\hspace{.1pt}\discretionary{.}{%
}{.}\hspace{.4pt}1467\discretionary{%
}{-}{-}8659\hspace{.1pt}\discretionary{.}{%
}{.}\hspace{.4pt}2011\hspace{.1pt}\discretionary{.}{%
}{.}\hspace{.4pt}01898\hspace{.1pt}\discretionary{.}{%
}{.}\hspace{.4pt}x}}


\bibitem{yang_many-to-many_2017}
Y.~Yang, T.~Dwyer, S.~Goodwin, and K.~Marriott.
\newblock Many-to-{{Many Geographically}}-{{Embedded Flow Visualisation}}: {{An
  Evaluation}}.
\newblock {\em IEEE Transactions on Visualization and Computer Graphics},
  23(1):411--420, Jan. 2017. doi: {{%
10\hspace{.1pt}\discretionary{.}{%
}{.}\hspace{.4pt}1109\discretionary{/}{%
}{/}TVCG\hspace{.1pt}\discretionary{.}{%
}{.}\hspace{.4pt}2016\hspace{.1pt}\discretionary{.}{%
}{.}\hspace{.4pt}2598885}}


\bibitem{yee_animated_2001}
K.-P. Yee, D.~Fisher, R.~Dhamija, and M.~Hearst.
\newblock Animated exploration of dynamic graphs with radial layout.
\newblock In {\em {{IEEE Symposium}} on {{Information Visualization}}, 2001.
  {{INFOVIS}} 2001.}, pp. 43--50, 2001. doi: {{%
10\hspace{.1pt}\discretionary{.}{%
}{.}\hspace{.4pt}1109\discretionary{/}{%
}{/}INFVIS\hspace{.1pt}\discretionary{.}{%
}{.}\hspace{.4pt}2001\hspace{.1pt}\discretionary{.}{%
}{.}\hspace{.4pt}963279}}


\end{thebibliography}
\end{document}